\documentclass[9pt,twocolumn,nofootinbib,superscriptaddress,longbibliography]{revtex4-1}
\usepackage{graphicx}
\usepackage{dcolumn}
\usepackage{bm,bbm}
\usepackage{xcolor}
\usepackage{tcolorbox}
\usepackage{algorithm}
\usepackage{algpseudocode}
\usepackage{amsmath}
\usepackage{bbold}
\usepackage{braket}
\usepackage{nameref}
\usepackage[toc,page]{appendix}

\usepackage[colorlinks,breaklinks,linkcolor={blue},citecolor={magenta},urlcolor={blue}]{hyperref}
\usepackage{enumerate}%allows different styles of enumerate environment

\usepackage[utf8]{inputenc}
\usepackage[T1]{fontenc}
\usepackage{lineno,commath,svg,multirow,fancyhdr,physics,float,lipsum,capt-of,verbatim,matlab-prettifier,epstopdf,latexsym,mathtools,color}
\begin{document}
\title{Beam deflection and negative drag in a moving nonlinear medium}

\author{Ryan Hogan}
\email{rhoga054@uottawa.ca}
\affiliation{Department of Physics, University of Ottawa, Ottawa, ON K1N 6N5, Canada}

\author{Akbar Safari}
\affiliation{Department of Physics, University of Ottawa, Ottawa, ON K1N 6N5, Canada}

\author{Giulia Marcucci}
\affiliation{Department of Physics, University of Ottawa, Ottawa, ON K1N 6N5, Canada}

\author{Boris Braverman}
\affiliation{Department of Physics, University of Ottawa, Ottawa, ON K1N 6N5, Canada}

\author{Robert W. Boyd}
\affiliation{Department of Physics, University of Ottawa, Ottawa, ON K1N 6N5, Canada}
\affiliation{Institute of Optics, University of Rochester, Rochester, NY 14627, USA}

\begin{abstract}
Light propagating in a moving medium with refractive index other than unity is subject to light drag. While the light drag effect due to the linear refractive index is often negligibly small, it can be enhanced in materials with a large group index. Here we show that the nonlinear refractive index can also play a crucial role in propagation of light in moving media and results in a beam deflection that might be confused with the transverse drag effect. We perform an experiment with a rotating ruby crystal which exhibits a very large negative group index and a positive nonlinear refractive index. The negative group index drags the light opposite to the motion of the medium. However, the positive nonlinear refractive index deflects the beam towards the motion of the medium and hinders the observation of the negative drag effect. Hence, we show that it is necessary to measure not only the transverse shift of the beam, but also its output angle to discriminate the light-drag effect from beam deflection --- a crucial step missing in earlier experiments.
\end{abstract}
\date{\today}
\maketitle
% \setboolean{displaycopyright}{true}

% \maketitle

\section*{Introduction}
\begin{figure}[t!]    %Fig1
\includegraphics[width=.99\linewidth]{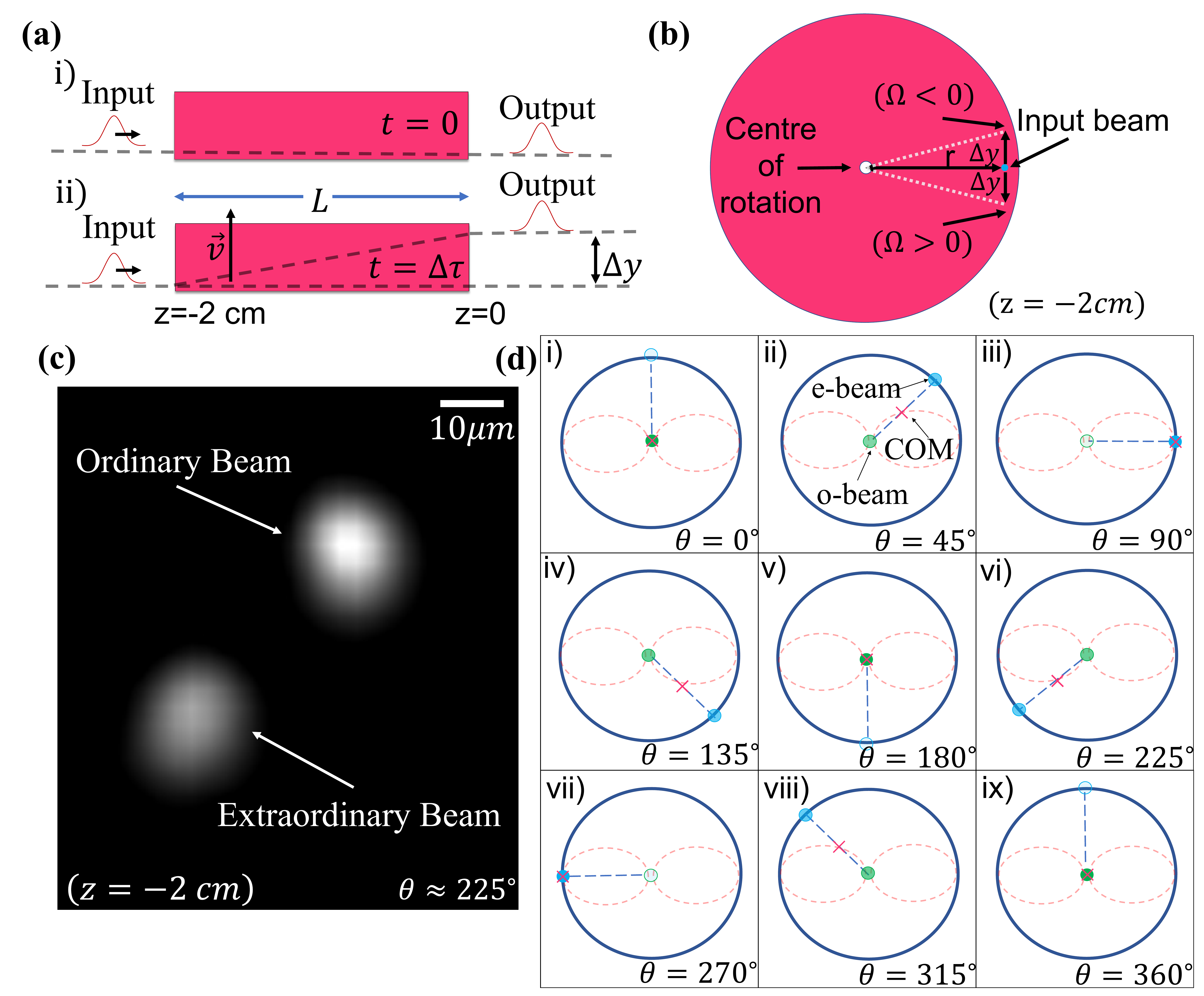} 
\caption{\textbf{(a)}  A schematic showcasing the beam being transversely shifted in a moving versus stationary medium. Although we use a CW laser, for simplicity of illustration we show the laser as pulses. Cases i) and ii) show the propagation of the laser in a stationary and moving medium, respectively. \textbf{(b)} The beam is far away from the center of rotation, which approximates linear motion in the y direction. The beam moves in the -y (+y) direction when the crystal rotates clockwise (counterclockwise) direction. \textbf{(c)} A single frame imaged at the front face of the crystal ($z=-2$ cm) that shows the two output beams, being the o- and e-beams due to the birefringence that propagate through the 2-cm-long ruby crystal. \textbf{(d)} A diagram showing the trajectories of o- and e- beams at different crystal orientations highlighting the change in intensity of each beam at 45-degree intervals. The red "x" shows the center of mass position for different crystal orientations highlighting the emergence of a figure-eight-like pattern, while o- and e- beams are shown by green and blue dots, respectively, with varying transparency to signify their relative intensities.}
\label{Figure0} 
\end{figure}

Propagation of light in moving media has been studied for more than two centuries \cite{fresnel1818,fizeau1860xxxii, balazs1955propagation, parks1974fresnel, leonhardt2001slow, artoni2001fresnel, gotte2007dragging, PhysRevA.68.063819, leonhardt2000relativistic, khan2021fizeau, qin2020fast}. Upon propagation, the trajectory of light can be manipulated through self-action effects \cite{chiao1964self,fibich2000critical}, beam deflection \cite{meyer1990optical,dixon2009ultrasensitive}, photon drag \cite{gibson1970photon,yee1972theory,gibson1980photon} and many other phenomena. The photon drag effect was hypothesized by Fresnel \cite{fresnel1818}, and then experimentally observed by Fizeau \cite{fizeau1860xxxii}. Fizeau's landmark experiment measured the shift of interference fringes within an interferometer containing a tube with moving water. These shifts in the fringes supported the idea that light is dragged in moving media. This phenomenon has gained increasing interest in the field of optics and is indeed still investigated in modern day research \cite{grinberg1988theory,wieck1990observation,artoni2001fresnel,strekalov2004observation,gotte2007dragging,davuluri2012controllable,kazemi2018phase,khan2021fizeau}. Photon drag can be longitudinal or transverse, i.e., along or perpendicular to the light propagation direction respectively. This article focuses on transverse rotary photon drag \cite{franke2011rotary}, distinctly different than longitudinal drag, given by
\begin{equation}
\Delta y = \frac{v}{c} \left(n_g-\frac{1}{n_{\phi}}\right) L,
\label{equation1}
\end{equation}
with $v$ the speed of the medium, $c$ the speed of light in vacuum, $L$ the length of the medium, $n_g$ and $n_{\phi}$ the group and phase indices, respectively. Photon drag scales linearly with group index. Typically phase and group indices are not large, and therefore do not create large transverse shifts. Recent studies show larger shifts using slow light  media(i.e. large group indices) \cite{franke2011rotary,kazemi2018phase,qin2020fast,safari2016light}. Figure \ref{Figure0}a) sketches the light propagation in a medium of length $L$ in two cases, a) a stationary medium, and b) a medium moving transversely with speed $v$. Experimentally, rotation is more feasible than linear motion. The beam is incident on the medium at a distance $r$ from the center of rotation, and using a slow light medium with $n_g\gg1/n_{\phi}$, the transverse drag can be simplified to 
\begin{equation}
    \Delta y \approx n_g L \left(\frac{r\Omega}{c}\right),
\label{equation2}
\end{equation}
where $\Omega$ is the rotational speed of the medium. Note that the beam size is much smaller than the medium radius, $r$. 
Large group indices are often achieved by employing a nonlinear phenomena such as coherent population oscillations (CPO) and electromagnetically induced transparency (EIT), that produce $n_g= 10^6$ or even larger. However, as we show below, in the presence of a strong saturating beam, one must consider nonlinear deflection in a moving medium which can be larger than and confused with the photon-drag effect. 
In a nonlinear medium, the impinging light can saturate the transition and locally change the refractive index of the medium. When the response of the medium is not instantaneous, as the medium moves in the transverse direction, the imprinted refractive index profile is dragged along with the motion of the medium Therefore, in a moving nonlinear medium, location of peak index change is shifted with respect to the center of the impinging light. Thus, the light sees a gradient in the refractive index and deflects at an angle. The sign of this nonlinear deflection depends on the sign of the nonlinear refractive index and the direction of motion of the medium. In a typical nonlinear interaction with positive nonlinear refractive index, where self-focusing is observed, the beam deflects towards the motion of the medium and thus resembles a positive photon-drag effect. In nonlinear deflection the output beam leaves the moving medium at an angle with respect to the input beam, while in the photon-drag effect the output beam is in parallel to the input beam. Therefore, one can distinguish the nonlinear deflection from the drag effect by measuring the output angle of the beam.
While the enhanced photon-drag effect depends on the group index including any nonlinear contribution (see Eq. (\ref{equation2})), the nonlinear deflection depends on the nonlinear refractive index of the medium. Thus, it seems that one should be able to achieve a large enhancement in the drag effect with negligible nonlinear deflection. However, according to the Kramers-Kronig relation, a large group index often is associated with a sluggish response~\cite{toll1956causality}. Therefore, if the large group index is achieved through a nonlinear interaction, one has to be careful with the nonlinear deflection and measure the output angle, a critical step missing in previous works \cite{franke2011rotary,jones1972fresnel,leach2008aether}.
{In this article, we use a rotary ruby rod to study the nonlinear light propagation in a moving medium. In a similar fashion to alexandrite \cite{bigelow2003superluminal}, ruby exhibits a large negative group index ($n_g \approx -10^6$) at wavelength 473 nm\cite{safari2022group}. Hence, according to Eq. (\ref{equation2}) one expects to observe a large negative photon-drag effect in which the position of the beam shifts in the direction opposite to the motion of the medium. }
\begin{figure}[t!]
\centering
\includegraphics[width=0.99\linewidth]{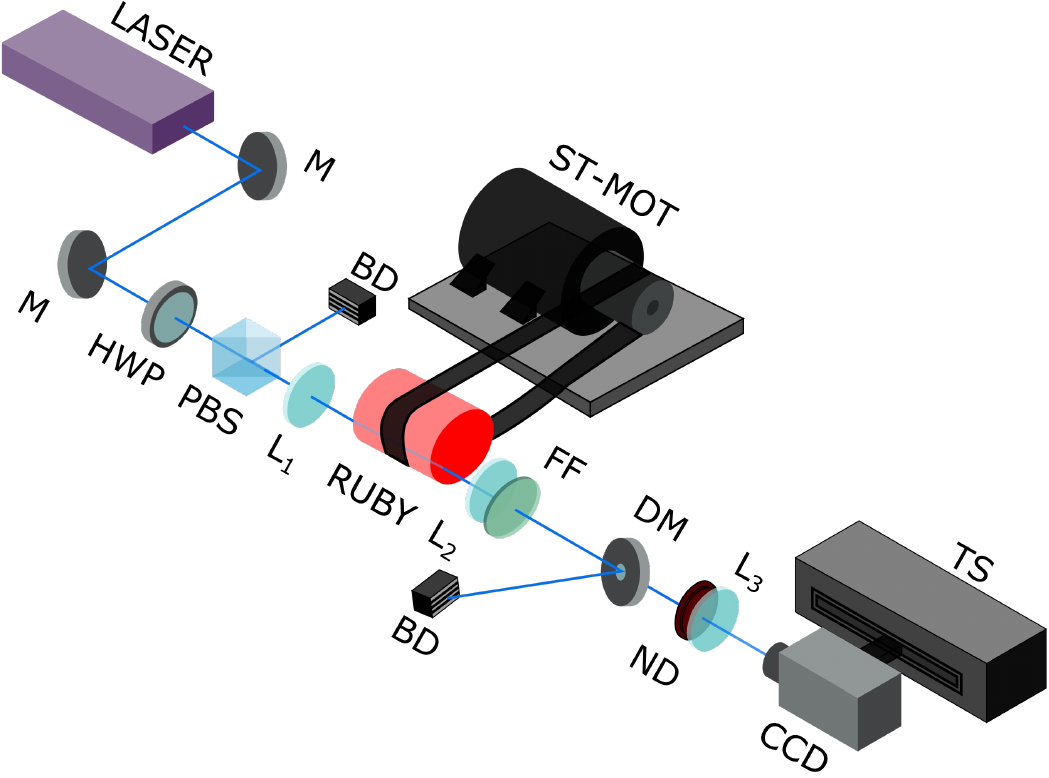}
\caption{A 520 mW continuous-wave laser beam at 473 nm is focused using a 100 mm focal length plano-convex lens $L_1$ to a spot size of 20 $\mu$m onto the input face of rotating ruby rod. The rod spins around its axis driven by a stepper motor. The laser beam at the output of the crystal is imaged onto a CCD camera with unity magnification using a 4-f system consisting of two lenses $L_2$ and $L_3$ of focal length $f=150$ mm. The CCD camera captures the beam, with a frame rate of 1000 fps, as the stepper motor is rotated at various speeds. An ND filter is placed between the dielectric mirror and lens 2, $L_2$ for nonlinear measurements, and between $L_1$ and the ruby for linear measurements. The CCD camera images at different z positions using a translation stage. Measurements are taken at $z=0$, $z=0.762 $ cm and $z=1.524$ cm to measure the transverse shift, as well as the output angle of the beam as it exits the crystal. The fluorescence filter $FF$ (high transmission near 473 nm) is used  to minimize fluorescence from the ruby rod from being collected by the CCD camera. The dielectric mirror $DM$ is used as a neutral density filter with low absorption to limit the beam intensity for high-power tests, while also minimizing image distortions due to aberrations induced by thermal nonlinearities in a standard neutral density filter. Input beam power was controlled by a half-wave plate and polarizing beam-splitter before the ruby crystal. (M: Mirror, HWP: Half-wave plate, PBS: Polarizing beam-splitter, BD: Beam dump, $L_1$: Plano-convex lens [f = 100 mm], $L_2$: Plano-convex lens [f = 150 mm], $L_3$: Plano-convex lens [f = 150 mm], FF: Fluorescence filter, DM: Dielectric mirror, ND: Neutral density filter [O.D. 1], and a CCD: Charge-coupled device.)}
\label{Figure1} 
\end{figure}

\begin{figure*}
  \includegraphics[width=0.49\linewidth]{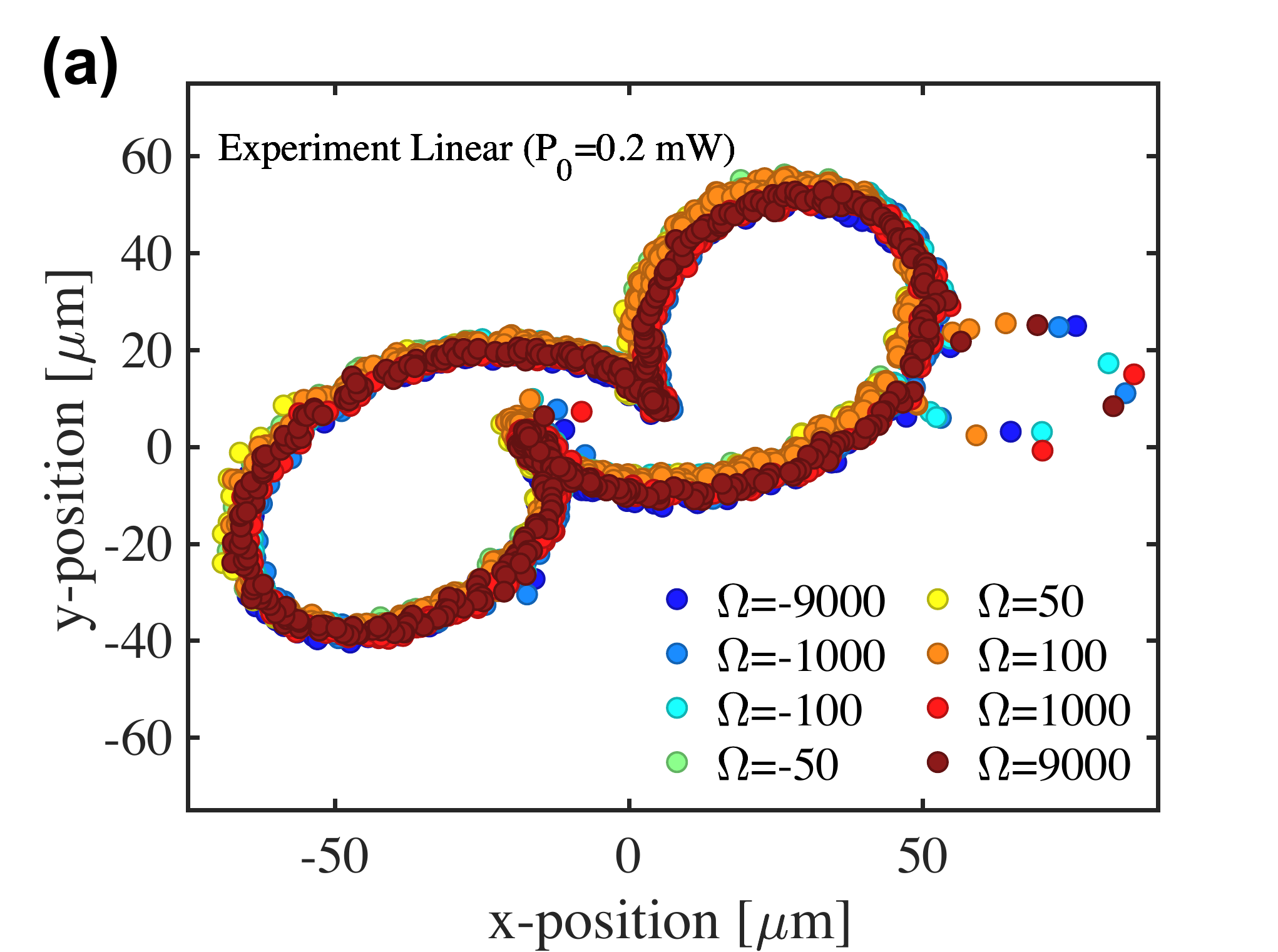}
  \includegraphics[width=0.49\linewidth]{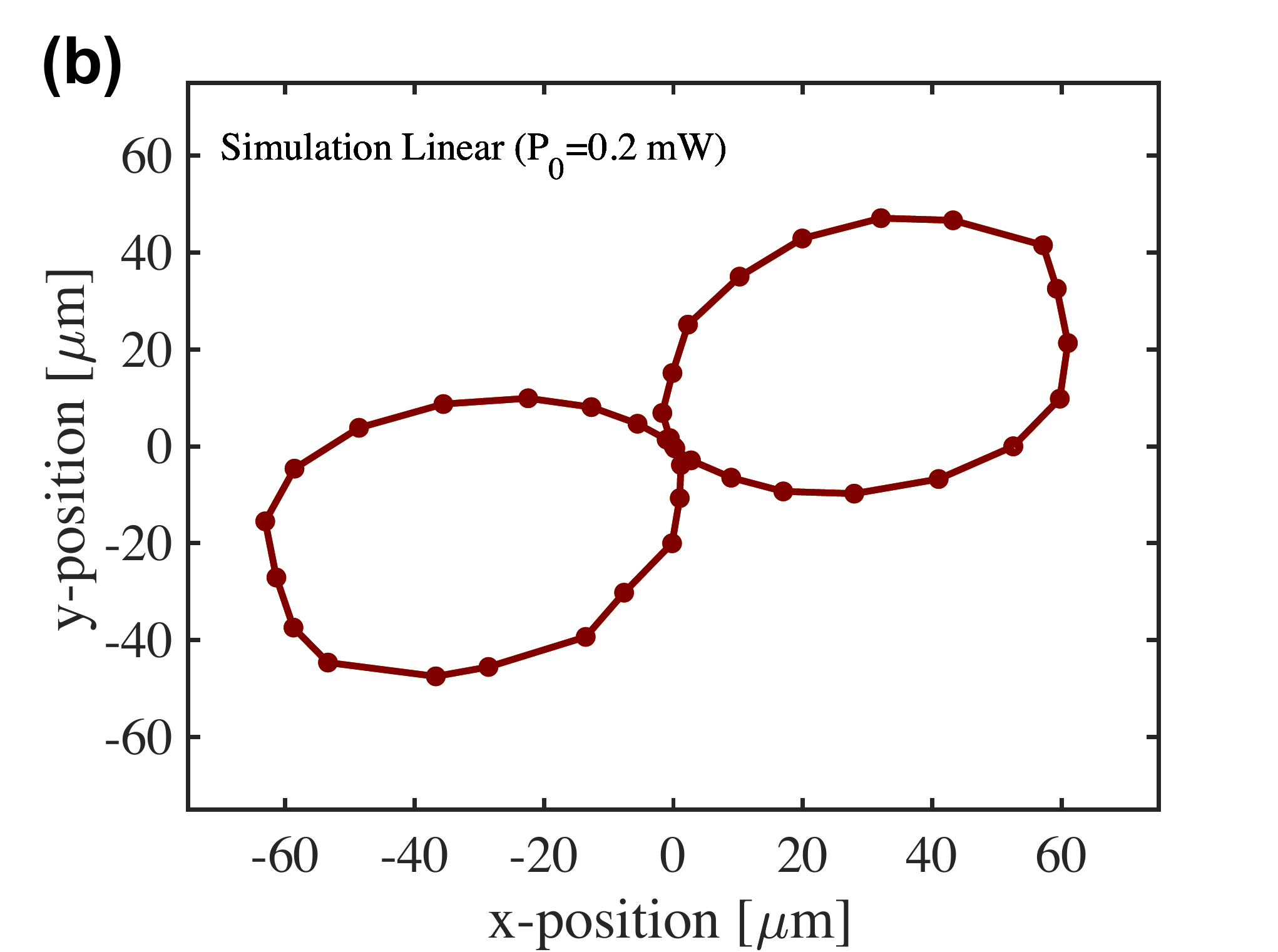} 
\caption{\textbf{(a)} Experimentally measured center of mass (COM) trajectories in the linear regime. \textbf{(b)} Simulated COM trajectories in the linear regime. Color scheme in the legend inset of the experimental data in \textbf{(a)} correspond equivalently the rotation speeds ($\Omega$) in units of degs/s in the simulated data. Trajectories of COM of the o- and e- beams (schematic shown in Figure \ref{Figure0} d) are plotted for an input laser power of 0.2 mW, considered as the linear regime. COM trajectories are plotted for rotation speeds of $\Omega = \pm 50, \pm 100, \pm 1000$, and  $\pm 9000$ deg/s. Here,  clockwise and counterclockwise rotation (looking into the beam) correspond to positive and negative rotation speeds, respectively. The COM for each speed follows the same figure-eight-like trajectory, since the intensity is too low to introduce deviations in the transverse movement due to nonlinearity or photon drag. The figure-eight-like pattern does not close in the center for the experimental results due to the polarization impurity in the low power regime.}
\label{Figure3}
\end{figure*}

Nevertheless, since ruby also exhibits nonlinear refraction, the beam deflects towards the direction motion of the medium due to nonlinear deflection. 
Because of the birefringence of the crystal, the input beam splits into ordinary (o) and extra-ordinary (e) beams which separate upon propagation in the crystal. The e-beam revolves with the rotation of the ruby rod. Moreover, the propagation of the o- and e-beam are coupled through the nonlinear interaction in ruby which creates an attractive force between the beams and further complicates their trajectory. We study this trajectory experimentally and simulate the propagation using nonlinear Schrodinger equations. Due to the simultaneous presence of birefringence, intensity-dependent photon drag, and strong nonlinearity, ruby can serve as a solid-state platform rich in physics with potential applications to beam steering \cite{boccia2009tunable,golub1992beam}, polarization detection \cite{allen1963new,gong2010review}, image rotation\cite{leach2008aether,franke2011rotary} , and potential for solitonic behaviour with associated applications \cite{hasegawa2000historical,ablowitz2000optical,kivshar1998dark}.
\section*{Methods}
The laser source used in the experiment, as shown in Fig. \ref{Figure1}, is a continuous-wave (CW) diode-pumped solid-state laser operating at 473 nm with an output power of 520 mW. We control the power of the laser beam using half-wave plate and polarizing-beam splitter. We use a 2-cm-long ruby rod, 9 mm in diameter, with a Cr$^{3+}$ doping concentration of 5\%. We focus the laser beam onto the front face of the crystal using a plano-convex lens of focal length $f=100$ mm, resulting in a 20 $\mu$m beam diameter located near the edge (0.1 mm away) of the ruby crystal face far from the center of rotation. The ruby was mounted in a hollow spindle whose rotation was controlled by a stepper motor and belt. The back face of the crystal was imaged onto a CCD camera using a 4-f lens system.

Shining linearly polarized light onto the rotary birefringent medium, the light sees two refractive indices upon propagation, $n_o=1.770$, and $n_e=1.762$, respectively. Without any influence of nonlinearity or photon drag, the two beams (o- and e-beams) then propagate with a finite angle separation of $\gamma_b= 8$ mrad, known as birefringent walk-off. The relative beam intensity reaches maxima and minima each quarter turn of the crystal (i.e., $\Delta \theta =90^{\circ}$). The beam input is aligned such that, regardless of crystal orientation, the o-beam propagates directly through the crystal, while the e-beam revolves about the o-beam. We track the motion of the average position of these two beams with an approach using the center of mass (COM), represented as a red “x” in Fig. \ref{Figure0}d. The COM is representing the centre of intensity distribution. This method is used since transverse beam profiles become larger on propagation and begin to overlap. Figure \ref{Figure0}c) shows a distinct Gaussian beam shape at $z=-2$ cm. In contrast, the o and e beams are mostly overlapped at the crystal back face, $z=0$ cm where transverse shifts are measured.

\begin{figure*}
  \includegraphics[width=0.49\linewidth]{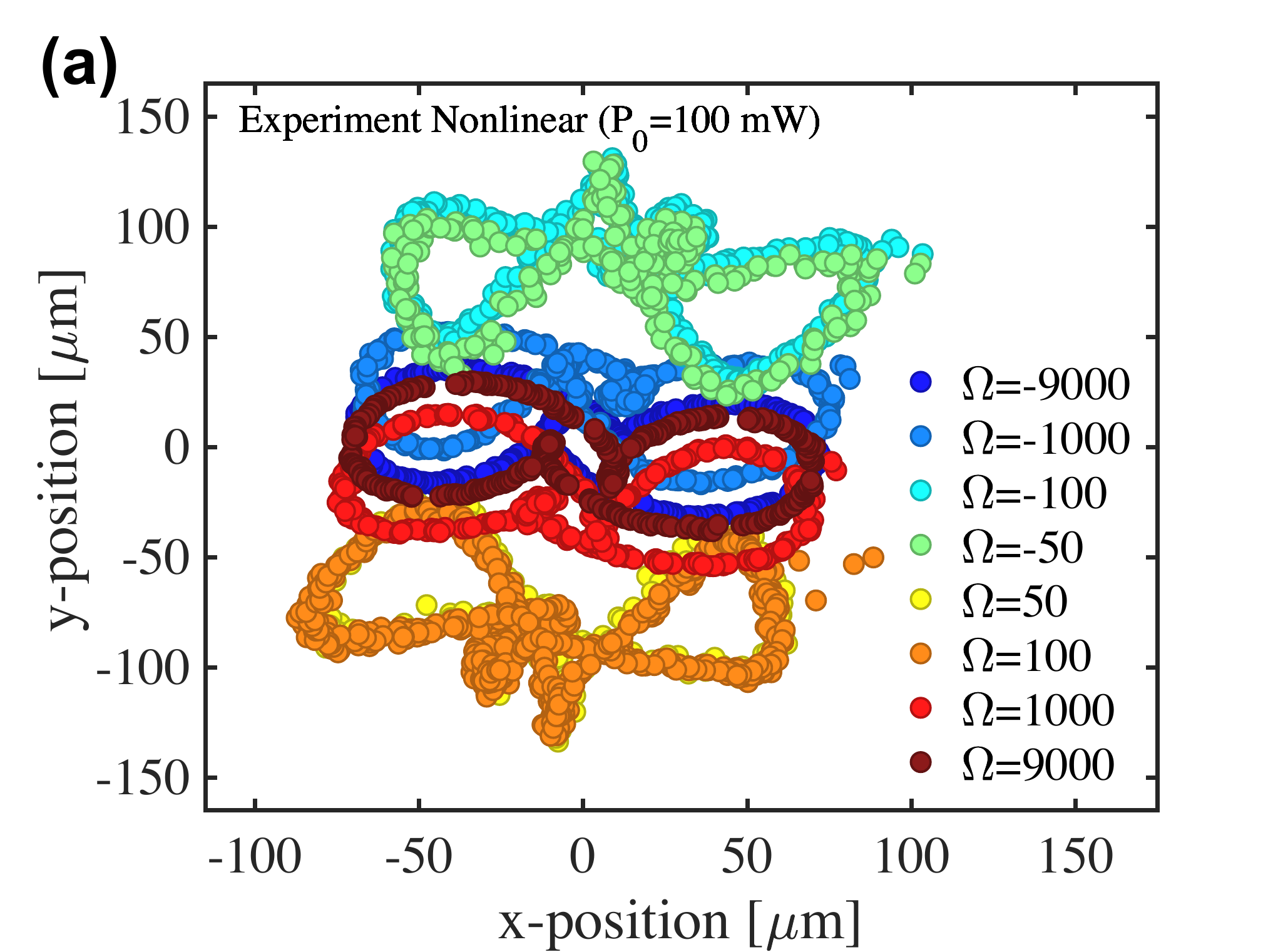}
  \includegraphics[width=0.49\linewidth]{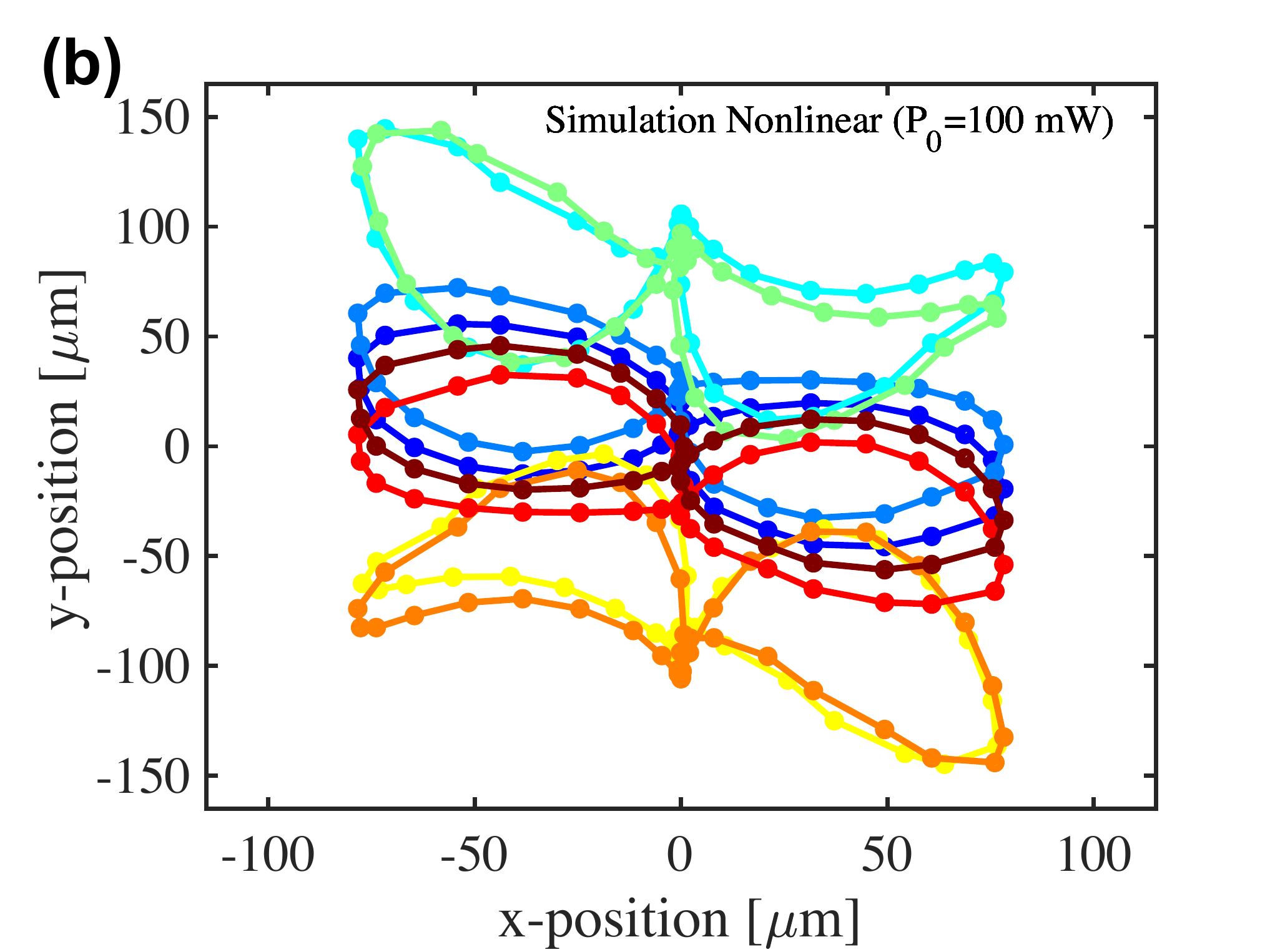}
\caption{\textbf{(a)} Experimentally measured center of mass (COM) trajectories in the nonlinear regime. \textbf{(b)} Corresponding simulated COM trajectories. COM trajectories of the o- and e- beams (schematic shown in Figure \ref{Figure0} d)) are plotted for an input laser power of 100 mW, considered as the nonlinear regime for different rotational speeds ($\Omega$) in units of degs/s. Here, clockwise and counterclockwise rotation correspond to positive and negative rotation speeds, respectively. At low speeds, o- and e- beams couple to each other causing significant variation in the traces of the COM upon rotation. Therefore, these patterns are different from the figure-eight-like patterns seen in the linear regime. At high speeds, the deviations from a figure-eight pattern start to average out. All that remains in the high-speed limit is that the figure-eight patterns are shifted from one another for positive and negative rotation speeds as a result of nonlinear deflection.}
\label{Figure4}
\end{figure*}

\section*{Results}
We measure the COM at $z=0$ for rotational speeds of $\Omega = \pm 50, \pm 100, \pm 1000$, and  $\pm 9000$ deg/s in clockwise (positive) and counterclockwise (negative) directions at three different input powers of 0.2 mW, 100 mW, and 520 mW, corresponding to weak, moderate, and intense illumination, respectively. We plot the COM trajectories for an input laser power of 0.2 mW, considered as the linear regime in Fig. \ref{Figure3}. We observe that all speeds trace out figure-eights and do not drift transversely.

\begin{figure*}
  \includegraphics[width=0.49\linewidth]{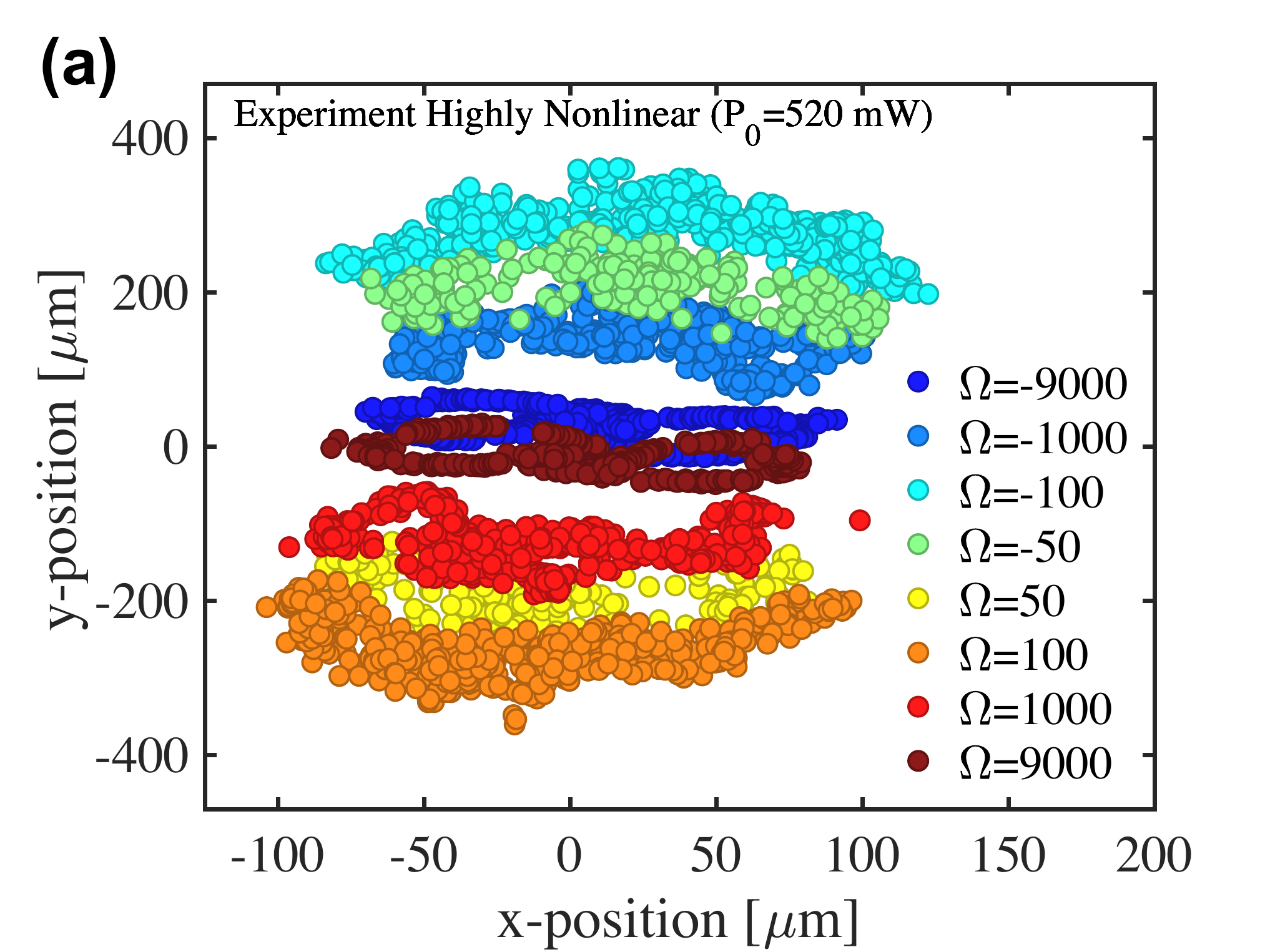}
  \includegraphics[width=0.49\linewidth]{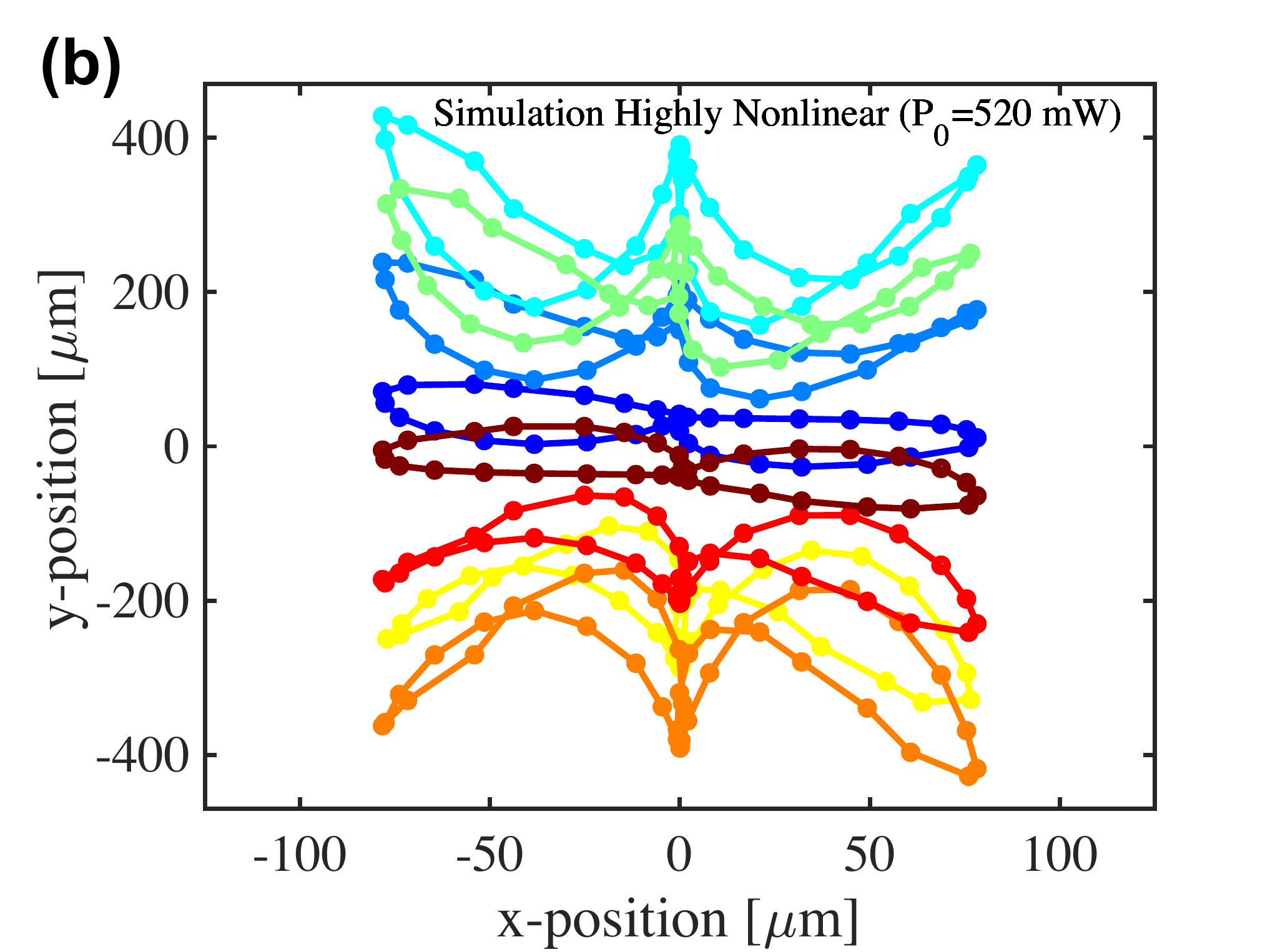}
\caption{\textbf{(a)} Experimentally measured COM trajectories in the highly nonlinear regime. \textbf{(b)} Corresponding simulated COM trajectories. Trajectories of COM of the o- and e- beams (schematic in Figure \ref{Figure0} d) are plotted for an input laser power of 520 mW, considered as the highly nonlinear regime for different rotational speeds ($\Omega$) in units of degs/s. Here,  clockwise and counterclockwise rotation correspond to positive and negative rotation speeds, respectively. At low speeds, trajectories are significantly distorted and have similar paths to the 100 mW results, but with more distortion due to stronger nonlinear coupling between the beams. At high speeds, the coupling between the beams is weaker due to the finite response time of the medium. For slow speeds $\Omega \leq100$, the trajectories are very noisy, and no discernable pattern is easily observed. This behaviour is mainly due to the thermal gradient impressed on the crystal by the intense illumination and therefore the transverse beam shape is drastically modified.}
\label{Figure5}
\end{figure*}

Figures \ref{Figure4} and \ref{Figure5} show COM trajectories in nonlinear and highly nonlinear regimes. At low speeds ($\Omega \leq 100$ deg/s), the o- and e- beams couple to each other causing significant variation in the traces of the COM upon rotation. Increasing intensity increases the thermal gradient impressed on the crystal drastically modifying the transverse beam shape. At high speeds, the deviations from a figure-eight-like pattern start to average out for input power of $P_0 = 100 $ mW and resemble those of the linear results, with the trajectories transversely shifted from one another based on the rotation speed. This is further seen in the highly nonlinear regime $P_0 = 520 $ mW, but at a slower pace due to more noise.  With lower speeds, one notes that the figure-eight-like COM trajectories knot near the center as a result of the nonlinear coupling of the o- and e- beams. Simulations are compared showing agreement in the traced patterns, and magnitudes of transverse shift, to be discussed in more detail later in this article. 

We extract the average position of these COM trajectories over an integer number of full rotations. Figure \ref{Figure6} shows the rotation speed dependence of the extracted transverse shift  at $z=0$ for linear, nonlinear and highly nonlinear regimes. The linear regime ($P = 0.2$ mW) shows a shift of a few microns. There is no clear scaling with rotation in the linear regime, however nonlinear regimes indeed show a trend similar to that of a log-normal distribution centered around $\Omega = 100$ deg/s. Nonlinear regimes show significant increases in the magnitude of transverse shift ($\Delta y(\Omega, I)$), showing that the observed shift is mainly nonlinear.
Transverse spatial shift is comprised of the nonlinear photon drag and nonlinear deflection. While the photon-drag shifts the beam in the transverse direction in parallel to the input beam, the nonlinear deflection deflects the beam at an angle. Thus, we distinguish the effects by measuring the transverse position at $z=0$ as well as at two locations after the crystal to find the output angle (see Fig. \ref{Figure7}a). This angle is calculated from the difference in average transverse position of the COM at different z positions, shown in Fig. \ref{Figure7}b. As expected from nonlinear deflection, this angle is intensity- and rotation speed- dependent. Therefore, the measurement of the angle confirms that the transverse shift observed at the end of the crystal is not only a result of the photon drag effect, but also nonlinear deflection. It is important to note that by moving the camera closer to the crystal one can image the input face of the crystal at which a seemingly negative drag is observed, i.e. the beams appear to be shifted opposite directions to the motion of the crystal. However, we highlight that such a measurement simply extrapolates the output beams towards the input face of the crystal. In reality, one images through a nonlinear medium, which is overlooked by the extrapolation. Indeed, to understand the position at the crystal input face, one must take care of the nonlinear effects on imaging to know the position accurately. Nonetheless, the extrapolation leads to a seemingly negative drag effect as a consequence of the large output angle due to nonlinear deflection. Therefore, in any measurement of the transverse drag effect, it is crucial to measure the output angle to obtain an accurate result.

\section*{Discussion}
\subsection*{A. Nonlinear refraction}
Intense linearly polarized light in rotating birefringent medium causes o- and e-beams both experience nonlinear refraction as the maximum intensity continuously moves between them creating a moving index gradient. The gradient leads to nonlinear coupling between the beams, where the local index variation pulls one beam toward the other with the higher refractive index, locally distorting the figure-eight-like COM trajectory. The distortions are dictated by the rotation speed, where the speed controls the amount of time that beam imprints an index gradient on the crystal.
 
The beam coupling increases in strength up to a  with decreasing rotation speed. The strongest coupling is seen at low speeds when the beams have sufficient time to imprint the maximum nonlinear index. On the other hand, higher rotation speeds imprint less gradient, blurring the effect of nonlinear refraction and non-distorted figure-eight-like trajectories are recovered. 

The regimes of rotations speeds are following sampling timescales based on two different origins: 1) Optical (slow) and 2) Thermal (fast). Often optical timescales are faster than thermal, but the optical response is attributed to CPO \cite{franke2011rotary} is on the order of 3-5 ms which is proportional to the faster rotation speeds. As for slow rotation speeds,  these sampling timescales are like thermal nonlinearities on the order of several hundred $\mu$s.

We model the temporal dynamics in these two regimes using a phenomenological fit consisting of two decaying exponentials, to be discussed later in this article, where we take analogy to spatial self-steepening \cite{de1992self,hernandez2020soliton}. That is, the beam is shifted due to the rotation speed acting on the group index, and therefore the group velocity. The rotation speed samples the dynamics representing a non-instantaneous temporal response of the system. The index gradient due to nonlinear refraction is discussed in the supplementary materials and how this is applied to a stationary medium in steady-state forming a Townes profile.

\subsection*{B. Simulations}
To better understand the experimental results, we model and simulate nonlinear propagation of linearly polarized light through a 2-cm-long rotating birefringent ruby rod, where o- and e-beams are created and vary in relative intensity upon rotation. Due to the weak birefringence typically associated with ruby and intense illumination, these two beams couple to each other upon rotation. Both beams are modelled using the nonlinear Schrödinger equation where we apply a split-step Fourier method and propagate the two with a coupling term containing a nonlinear response function. Following the derivation of Marcucci \textit{et al.} \cite{marcucci2019optical}, we write wave equations for the medium using a Kerr-type nonlinearity of thermal origin. Rotation and birefringence are also included \cite{conti2005spatial}. Furthermore, a term for the effective group index is incorporated into the coupled equations which is intensity- and rotation speed-dependent. 

\begin{figure}[t!] 
\centering
\vspace{0cm}
\includegraphics[width=\linewidth]{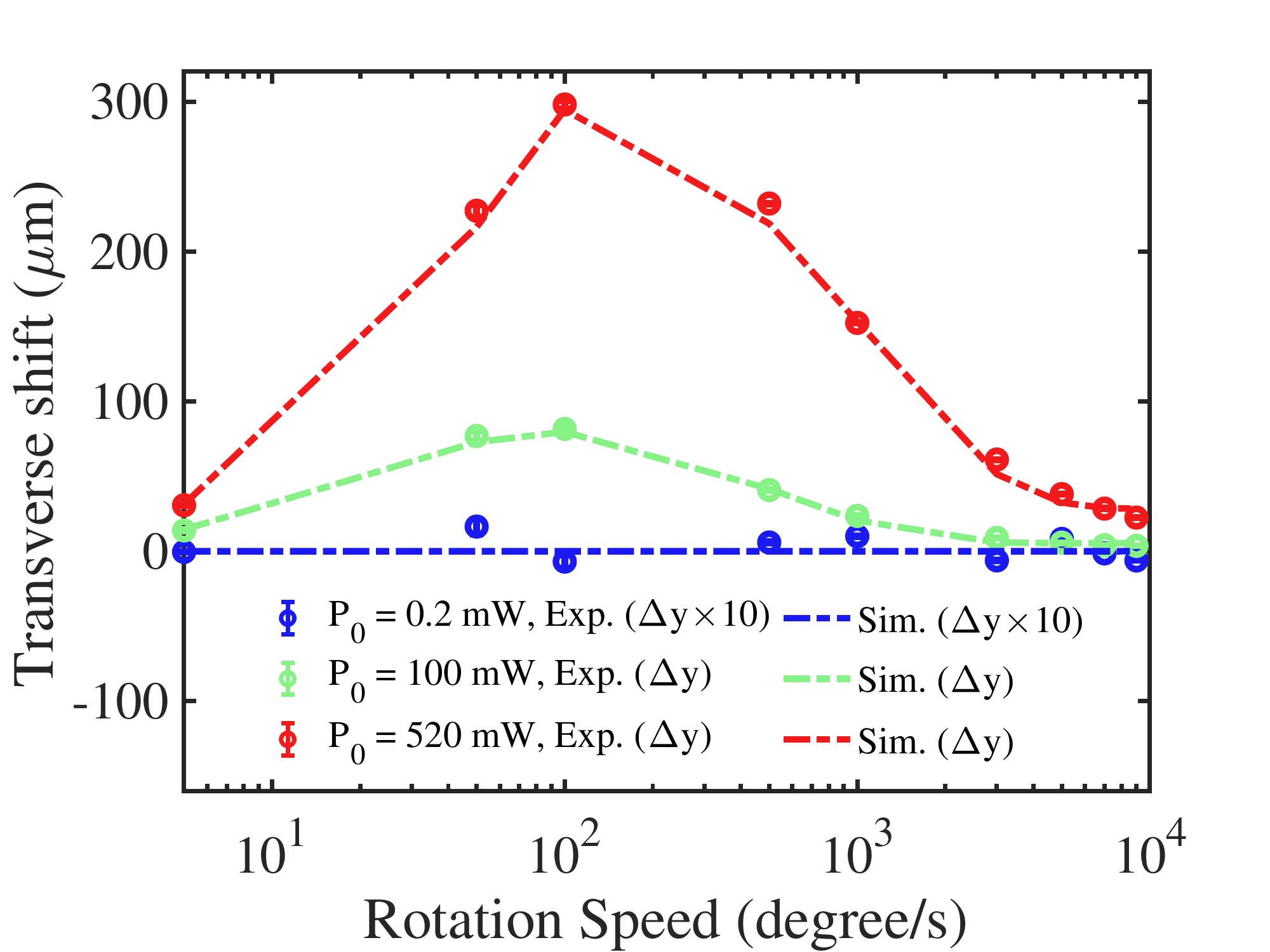} 
\caption{Experimental and simulated amount of shift in the beam's transverse position at the end of the crystal for 0.2 mW, 100 mW, and 520 mW beam input laser power. The measured shift for the linear regime (i.e., $P_0 = 0.2$ mW) for both experiment and simulations is multiplied by a factor of 10, showing that there is no discernible deviation from zero shift. Simulations are plotted using dotted lines in green and red for the nonlinear and highly nonlinear regimes for better comparison to experimental data. The fits were based on a phenomenological exponential function in Eq. (\ref{equation2}). The fit is not a perfect match due to the simulated nonlinear response of the material acting on the beams upon propagation through the crystal.} 
\label{Figure6} 
\vspace{0cm}
\end{figure}

Using our theoretical framework, including the phenomenological fit convoluted with nonlinear propagation, we were able to accurately simulate the amount of transverse shift, and the transverse movement of the center of mass of the beams observed in experiment. We develop a set of generalized coupled nonlinear Schrodinger equations, written as follows
\begin{equation}
\begin{aligned}
-\frac{\partial E_o}{\partial z} + \frac{i}{2 k_o} \nabla^2_{\perp} E_o + \left(n_g^{\mathrm{eff}} \frac{\partial E_o}{\partial y} + \frac{i k_o}{n_o} \Delta n_{NL} E_o\right) = 0, 
\end{aligned}
\label{equationGNLSE1}
\end{equation}
\begin{equation}
\begin{aligned}
&-\frac{\partial E_e}{\partial z} + \frac{i}{2 k_e\cos^2(\gamma)}\nabla^2_{\perp} E_e\\&+ \left(\frac{i k_e}{n_e\cos^2(\gamma)}\Delta n_{NL}E_e+n_g^{\mathrm{eff}} \frac{\partial E_e}{\partial y}\right)\\&+2 \tan(\gamma) \left[ \cos(\Omega t) \frac{\partial E_e}{\partial x} +\sin(\Omega t) \frac{\partial E_e}{\partial y}\right] = 0, 
\end{aligned}
\label{equationGNLSE2}
\end{equation}
where the fields $E_o$ and $E_e$ represent the o- and e- beams, $k_{o,e}$ are the o- and e-beam wave vectors, $n_{o,e}$ are the o- and e-beam refractive indices, respectively. Furthermore, $\gamma$ is the tilt angle, $\nabla^2_{\perp}=\frac{\partial^2}{\partial x^2}+\frac{\partial^2}{\partial y^2}$ is the transverse Laplacian operator, and $\Delta n_{NL}$ is the nonlocal Kerr-type nonlinearity that contains the coupling term in the kernel function \cite{marcucci2019optical}. One notes that  the nonlinear deflection term works on the derivative of the field rather than the field directly like that of nonlinear refraction. When adding dispersion to the nonlinear Schrodinger equation, one includes this effect as $n_g^{\mathrm{eff}}\frac{\partial E}{\partial t}c^{-1}$, which becomes $n_g^{\mathrm{eff}}\frac{\partial E}{\partial y}$ when considering a continuous-wave laser rather than pulses. This term will drive the nonlinear deflection that is measured at the crystal back face, where $n_g^{\mathrm{eff}}:=n_g^{\mathrm{eff}}(\Omega,I)$ as in Eq. (\ref{equation5}).

\begin{figure}[t!]
  \includegraphics[width=0.99\columnwidth]{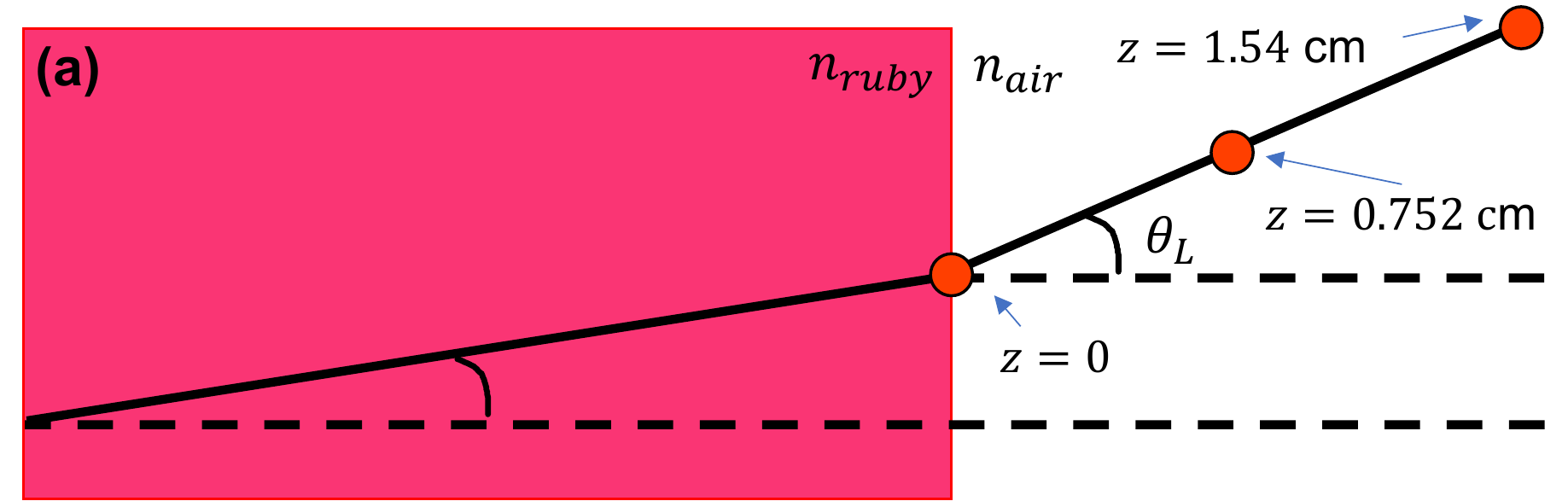}
  \includegraphics[width=0.99\columnwidth]{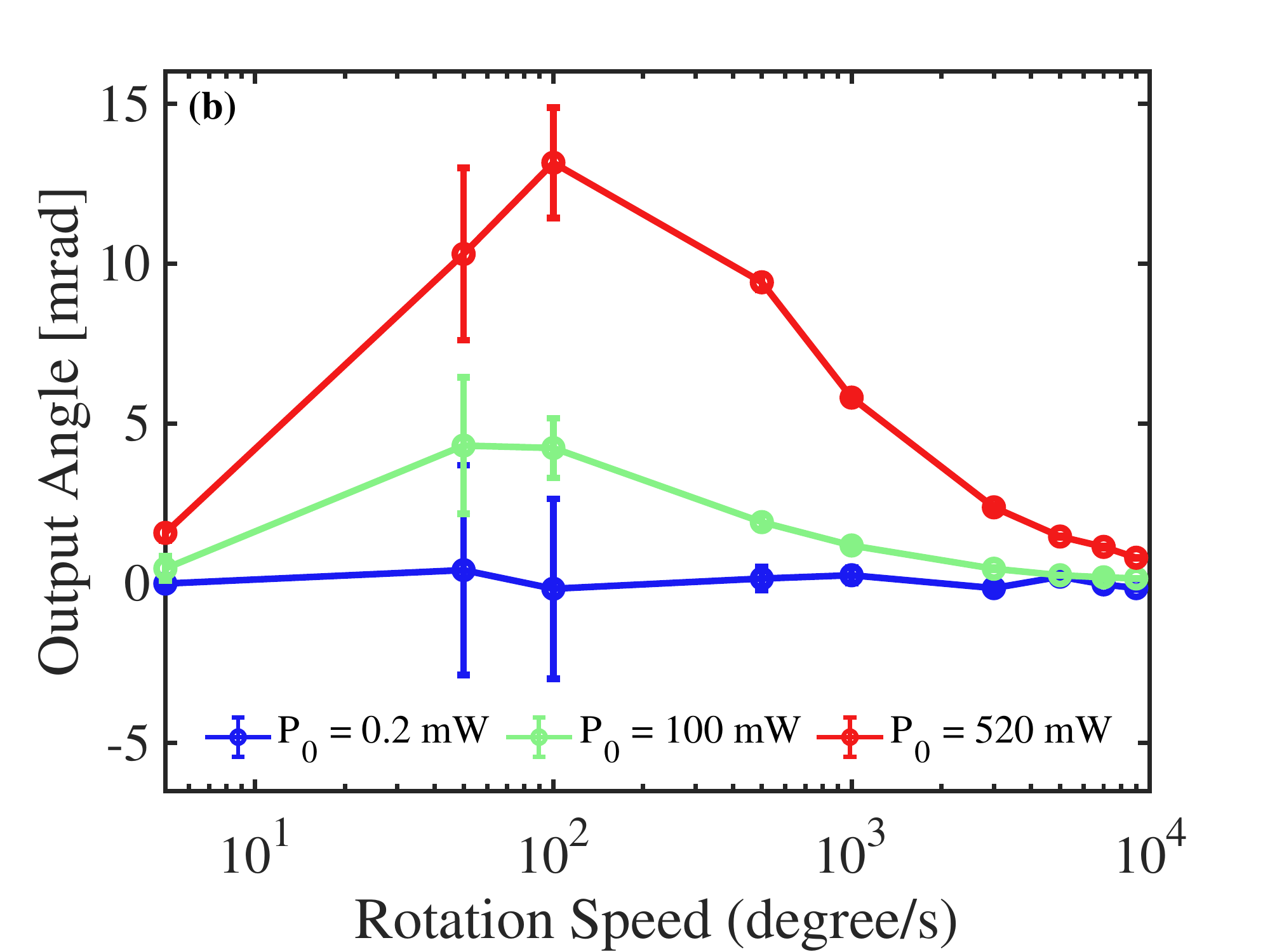}
\caption{\textbf{(a)} A schematic showing the output beam's angle after leaving the crystal. The nonlinear response of the crystal changes the angle at the interface of the crystal back face and therefore changes the propagation pathway. \textbf{(b)} The output angle and its uncertainty calculated from the beams' transverse positions measured at three points along the z-axis ($z=0$, $z=0.762$ cm and $z=1.524$ cm). As the laser's power increases, the output angle increases as expected from nonlinear deflection.}
\label{Figure7}
\end{figure}

The magnitude of the nonlinear deflection is proportional to the magnitude of the effective group index, controlled by the intensity and rotation speed. The rotation speed changes the conditions for how quickly the heat dissipates through the crystal, and thus the magnitude of the index gradient. If the speeds are sufficiently slow, the index gradient stays relatively constant and causes an increasing amount of transverse shift. Typically the time scale needed to deflect the beam is always very short (i.e. $2 \mathrm{cm}/(c/n_g)$, however once the maximum amount of transverse shift is met, i.e $\Omega\approx100$ deg/s, the crystal starts rotating faster than the timescale needed to form the index gradient. Typically,  Thus, as we increase the rotation speed, the beam sees less of an index gradient and therefore experiences less transverse shift. The curve associated  transverse shift versus rotation speed is comprised of two decaying exponentials centered about $\Omega = 100$ degs/s. The decay rates of these two exponentials gives rise to asymmetric distribution about $\Omega = 100$ degs/s. For slow speeds, the index gradient impressed on the crystal is not blurred leading to a slower decay rate. At higher speeds, the beam is sampling only some of the index gradient drastically changing the magnitude, and thus the decay rate is faster. This is shown in Fig. \ref{Figure6}, where the behaviour is not symmetric about $\Omega = 100$ deg/s.

 We draw analogy to a self-steepened pulse to explain the asymmetry of the transverse shift versus rotation speed. In a self-steepened pulse, the group velocity travels at different speeds dependent on the intensity. As such, different parts of the pulse travel at speeds according to the group velocity. Higher intensities are seen at the peak of the pulse that gives rise to large group index, and thus a slower group velocity. At the wings of the pulse, the intensity is lower and the group velocity is larger. This causes the pulse to become asymmetric in time, and thus the material response will have an asymmetric response in time. The rotation speed controls the amount of time that maximum intensity is in a given area, and therefore the index gradient will have asymmetric temporal response as well. As a result, the amount of transverse shift will also change asymmetrically.

A phenomenological fit for the transverse shift, $\Delta y$, was created using the experimental data in Fig. \ref{Figure6} which shows an exponential dependence between the rotation speed and the amount of transverse shift (see Supplementary Materials). The phenomenological fit has form of a decaying exponential, and thus we can represent the maximum imprinted nonlinear group index as $\Delta n_g=n_2^g I_{\mathrm{max}}(t)=n_2^g I_0 \exp(-t/\tau_c)$, where $I_0$ is the input intensity and $\tau_c$ is a characteristic time of the decay of the nonlinear response, which we ascribe to thermal diffusion as the dominant thermal contribution to the nonlinear response. This maximum group index gradient will change the size of the effective group index. Let us write the time in terms of the rotation speed as $t=\tau_c \Omega/\Omega_c$, where $\Omega$ is the rotation speed, $\Omega_c$ is a characteristic rotation speed, and we rewrite the transverse shift to be
\begin{equation}
\Delta y \approx \frac{ r\Omega L n_g^{\mathrm{eff}}}{c},
\end{equation}
where $r$ is the distance from the center of the crystal to the beam position, and $n_g^{\mathrm{eff}}$ is the effective group index written as
\begin{equation}
n_g^{\mathrm{eff}}= n_g^0 + n_2^g I = n_g^0+n_2^g I_o e^{-\Omega/\Omega_c}.
\label{equation4}
\end{equation}

We can look at the time-average response of the effective group index for a given speed. Therefore, if we break up the average temporal response into a fast and slow contribution, we can write the effective group index as
\begin{equation}
n_g^{\mathrm{eff}}= n_g^0+n_2^g I_o \left(\frac{1}{f_s} e^{-\Omega/\Omega_s}-f_f e^{-\Omega/\Omega_f}\right),
\label{equation5}
\end{equation}
where $n_2^g=10^7 $m$^2$/W, and $f_s$ and $f_f$ are scaling factors equal to 0.97 and 0.94, respectively. These values are similar to those when considering the peak power of a Gaussian pulse. $\Omega_s$ and $\Omega_f$ refer to the inverse time scales of the slow and fast parts of the interaction, where $\Omega_{s,f}=1/(2\pi \tau_{s,f})$. The slow and fast timescales are $\tau_s=3.5$ ms, and $\tau_f=175$ $\mu$s, respectively. These two timescales follow the timescale of the lifetime of the excited ions 3 to 5 ms \cite{franke2011rotary}, and the typical timescale of thermal diffusion ($\approx 200\mu$s) \cite{doiron2004time}. Equation (\ref{equation5}) can be seen plotted in the supplementary materials.

Fast and slow time scales modify the magnitude of the effective group index, representing an approximate non-instantaneous response. The nonlinear response of the medium is indeed non-instantaneous, but this approach to good approximation represents the dynamics in the system while alleviating computational expense when one includes a non-instantaneous response into simulations. Equation (\ref{equation5}) is then introduced into a generalized nonlinear Schrödinger equation, and nonlinear propagation is simulated to investigate the transverse COM trajectories, and extract the amount of transverse shift. We further incorporate other nonlinear effects on the output amount of transverse shift experienced by the beam COM. These results are shown in Figs. \ref{Figure3}, \ref{Figure4}, and \ref{Figure5}, achieving good agreement amongst the trajectories and the transverse shift due to photon drag and nonlinear deflection.

From Eq. \ref{equationGNLSE1} and \ref{equationGNLSE2}, we represent the static nonlinear refraction as an index gradient that takes the form
\begin{equation}
\begin{aligned}
&\Delta n_{NL}(x,y,\Omega t,\gamma) \\&= n_2 \iint \mathrm{d}\widetilde{x} \mathrm{d}\widetilde{y} K_\gamma \left(\Delta x, \Delta y, \Omega t \right) I(\widetilde{x},\widetilde{y})-n_{o,e}, 
\end{aligned}
\end{equation}
where $\Delta x=x-\widetilde{x}$, $\Delta y=y-\widetilde{y}$. Here, $\widetilde{x}$ and $\widetilde{y}$ are the Cartesian coordinates of an arbitrary position within the space where the nonlinear kernel function acts. The nonlinear potential in the laboratory's frame depends on the crystal's response function $K'$ \cite{marcucci2019optical}, which is given by the thermal properties of the material
\begin{equation}
\begin{aligned}
\begin{array}{lcl}
K_\gamma \left(x, y, \Omega t\right)/K' =\\ \left[\cos(\gamma) \cos(\Omega t) x+ \cos(\gamma)\sin(\Omega t) y, -\sin(\Omega t) x+ \cos(\Omega t) y\right],
\end{array}
\end{aligned}
\end{equation}
and $I=|E_o|^2+|E_e|^2$.

A further explanation of the theoretical modeling of light propagation through a moving nonlinear medium is discussed by Hogan \textit{et al.}\cite{hogan2022nlphoton}. Upon propagating two beams through the crystal, we extract two main parameters as a function of the input intensity, and rotation speed: 1) the position of the COM in the transverse plane, and 2) the transverse shift (overall average position of the COM trajectories) experienced by the beams at the back face of the crystal. The COM trajectories and the values of the transverse shift are determined for a variety of crystal rotation speeds, and the three powers used in experiment with the addition of the phenomenological fit in the drift term as a modified effective group index. Results of the simulations are presented in the Figs. \ref{Figure3}, \ref{Figure4}, and \ref{Figure5} for direct comparison to experimental data. 
\section*{Conclusion}
We have demonstrated experimentally and through simulation that a 2-cm long rotating ruby crystal illuminated with 473-nm light produces a transverse shift as a result of nonlinear photon drag and nonlinear deflection. In rotating saturable media with self-focusing nonlinear refraction, one must measure the output angle to distinguish nonlinear deflection and transverse photon drag. We note that even if the medium presents large negative group indices, nonlinear deflection can dominate over negative drag when nonlinear refraction is large and positive. The maximum transverse shift is found to be found to be $\Delta y = +300$  $\mu$m, and the maximum angular shift is found to be found to be $\theta= 13$ mrad at the back face of the crystal ($z = 0$). Moreover, exotic trajectories were observed experimentally for the center of mass of the beam in the transverse plane at the crystal back face, and reproduced in simulation with good agreement. Since the position of transverse profile of the beam is controllable by the rotation speed of the crystal and input intensity of the beam, one can imagine applications in beam-steering and image rotation, as well as understanding the resilience of state of polarization to the motion of the medium.
\vspace{-0.3 cm}
\section*{Funding}
This work was supported by the Canada Research Chairs program and the National Science and Engineering Research Council of Canada (NSERC). A.S. acknowledges the support of NSERC [funding reference number PDF - 546105 - 2020]. B.B. also acknowledges the support of the Banting postdoctoral fellowship of NSERC. R.W.B. also acknowledges support by the US Office of Naval Research.

\section*{Acknowledgments}
The authors would like to thank Xiaoqin Gao for valuable advice on making figures and structural formatting of the manuscript.

\section*{Disclosures}
The authors declare no conflicts of interest.

\section*{Data Availability Statement}
The reader can email the corresponding author, Ryan Hogan, for any code that may be needed for their purpose. He can be contacted at rhoga054@uottawa.ca. We will respond to reasonable requests for the code. 
\vspace{0.37cm}
\section*{Author contributions statement}
A.S. conceived the experiment, R.H. and A.S. conducted the experiment, R.H. and B.B. analysed the results, G.M. and R.H. developed theory and conducted simulations, R.H. wrote the article and R.B supervised this work. All authors reviewed the manuscript. 
\section*{Supplemental Document}
See Supplement 1 for supporting content. 
\bibliography{References_RDA2}

%merlin.mbs apsrev4-1.bst 2010-07-25 4.21a (PWD, AO, DPC) hacked
%Control: key (0)
%Control: author (0) dotless jnrlst
%Control: editor formatted (1) identically to author
%Control: production of article title (0) allowed
%Control: page (1) range
%Control: year (0) verbatim
%Control: production of eprint (0) enabled
\begin{thebibliography}{45}%
\makeatletter
\providecommand \@ifxundefined [1]{%
 \@ifx{#1\undefined}
}%
\providecommand \@ifnum [1]{%
 \ifnum #1\expandafter \@firstoftwo
 \else \expandafter \@secondoftwo
 \fi
}%
\providecommand \@ifx [1]{%
 \ifx #1\expandafter \@firstoftwo
 \else \expandafter \@secondoftwo
 \fi
}%
\providecommand \natexlab [1]{#1}%
\providecommand \enquote  [1]{``#1''}%
\providecommand \bibnamefont  [1]{#1}%
\providecommand \bibfnamefont [1]{#1}%
\providecommand \citenamefont [1]{#1}%
\providecommand \href@noop [0]{\@secondoftwo}%
\providecommand \href [0]{\begingroup \@sanitize@url \@href}%
\providecommand \@href[1]{\@@startlink{#1}\@@href}%
\providecommand \@@href[1]{\endgroup#1\@@endlink}%
\providecommand \@sanitize@url [0]{\catcode `\\12\catcode `\$12\catcode
  `\&12\catcode `\#12\catcode `\^12\catcode `\_12\catcode `\%12\relax}%
\providecommand \@@startlink[1]{}%
\providecommand \@@endlink[0]{}%
\providecommand \url  [0]{\begingroup\@sanitize@url \@url }%
\providecommand \@url [1]{\endgroup\@href {#1}{\urlprefix }}%
\providecommand \urlprefix  [0]{URL }%
\providecommand \Eprint [0]{\href }%
\providecommand \doibase [0]{http://dx.doi.org/}%
\providecommand \selectlanguage [0]{\@gobble}%
\providecommand \bibinfo  [0]{\@secondoftwo}%
\providecommand \bibfield  [0]{\@secondoftwo}%
\providecommand \translation [1]{[#1]}%
\providecommand \BibitemOpen [0]{}%
\providecommand \bibitemStop [0]{}%
\providecommand \bibitemNoStop [0]{.\EOS\space}%
\providecommand \EOS [0]{\spacefactor3000\relax}%
\providecommand \BibitemShut  [1]{\csname bibitem#1\endcsname}%
\let\auto@bib@innerbib\@empty
%</preamble>
\bibitem [{\citenamefont {Fresnel}(1818)}]{fresnel1818}%
  \BibitemOpen
  \bibfield  {author} {\bibinfo {author} {\bibfnamefont {A}~\bibnamefont
  {Fresnel}},\ }\bibfield  {title} {\enquote {\bibinfo {title} {Lettre
  d’augustin fresnel à françois arago sur l’influence du mouvement
  terrestre dans quelques phénomènes d’optique.}}\ }\href@noop {}
  {\bibfield  {journal} {\bibinfo  {journal} {Annales de Chimie et de
  Physique}\ }\textbf {\bibinfo {volume} {9}},\ \bibinfo {pages} {57--66; 286}
  (\bibinfo {year} {1818})}\BibitemShut {NoStop}%
\bibitem [{\citenamefont {Fizeau}(1860)}]{fizeau1860xxxii}%
  \BibitemOpen
  \bibfield  {author} {\bibinfo {author} {\bibfnamefont {MH}~\bibnamefont
  {Fizeau}},\ }\bibfield  {title} {\enquote {\bibinfo {title} {{XXXII.} {O}n
  the effect of the motion of a body upon the velocity with which it is
  traversed by light},}\ }\href@noop {} {\bibfield  {journal} {\bibinfo
  {journal} {The London, Edinburgh, and Dublin Philosophical Magazine and
  Journal of Science}\ }\textbf {\bibinfo {volume} {19}},\ \bibinfo {pages}
  {245--260} (\bibinfo {year} {1860})}\BibitemShut {NoStop}%
\bibitem [{\citenamefont {Balazs}(1955)}]{balazs1955propagation}%
  \BibitemOpen
  \bibfield  {author} {\bibinfo {author} {\bibfnamefont {Nandor~L}\
  \bibnamefont {Balazs}},\ }\bibfield  {title} {\enquote {\bibinfo {title} {The
  propagation of light rays in moving media},}\ }\href@noop {} {\bibfield
  {journal} {\bibinfo  {journal} {JOSA}\ }\textbf {\bibinfo {volume} {45}},\
  \bibinfo {pages} {63\_2--64} (\bibinfo {year} {1955})}\BibitemShut {NoStop}%
\bibitem [{\citenamefont {Parks}\ and\ \citenamefont
  {Dowell}(1974)}]{parks1974fresnel}%
  \BibitemOpen
  \bibfield  {author} {\bibinfo {author} {\bibfnamefont {WF}~\bibnamefont
  {Parks}}\ and\ \bibinfo {author} {\bibfnamefont {JT}~\bibnamefont {Dowell}},\
  }\bibfield  {title} {\enquote {\bibinfo {title} {Fresnel drag in uniformly
  moving media},}\ }\href@noop {} {\bibfield  {journal} {\bibinfo  {journal}
  {Physical Review A}\ }\textbf {\bibinfo {volume} {9}},\ \bibinfo {pages}
  {565} (\bibinfo {year} {1974})}\BibitemShut {NoStop}%
\bibitem [{\citenamefont {Leonhardt}\ and\ \citenamefont
  {Piwnicki}(2001)}]{leonhardt2001slow}%
  \BibitemOpen
  \bibfield  {author} {\bibinfo {author} {\bibfnamefont {U}~\bibnamefont
  {Leonhardt}}\ and\ \bibinfo {author} {\bibfnamefont {P}~\bibnamefont
  {Piwnicki}},\ }\bibfield  {title} {\enquote {\bibinfo {title} {Slow light in
  moving media},}\ }\href@noop {} {\bibfield  {journal} {\bibinfo  {journal}
  {Journal of Modern Optics}\ }\textbf {\bibinfo {volume} {48}},\ \bibinfo
  {pages} {977--988} (\bibinfo {year} {2001})}\BibitemShut {NoStop}%
\bibitem [{\citenamefont {Artoni}\ \emph {et~al.}(2001)\citenamefont {Artoni},
  \citenamefont {Carusotto}, \citenamefont {La~Rocca},\ and\ \citenamefont
  {Bassani}}]{artoni2001fresnel}%
  \BibitemOpen
  \bibfield  {author} {\bibinfo {author} {\bibfnamefont {M}~\bibnamefont
  {Artoni}}, \bibinfo {author} {\bibfnamefont {I}~\bibnamefont {Carusotto}},
  \bibinfo {author} {\bibfnamefont {GC}~\bibnamefont {La~Rocca}}, \ and\
  \bibinfo {author} {\bibfnamefont {F}~\bibnamefont {Bassani}},\ }\bibfield
  {title} {\enquote {\bibinfo {title} {Fresnel light drag in a coherently
  driven moving medium},}\ }\href@noop {} {\bibfield  {journal} {\bibinfo
  {journal} {Physical Review Letters}\ }\textbf {\bibinfo {volume} {86}},\
  \bibinfo {pages} {2549} (\bibinfo {year} {2001})}\BibitemShut {NoStop}%
\bibitem [{\citenamefont {G{\"o}tte}\ \emph {et~al.}(2007)\citenamefont
  {G{\"o}tte}, \citenamefont {Barnett},\ and\ \citenamefont
  {Padgett}}]{gotte2007dragging}%
  \BibitemOpen
  \bibfield  {author} {\bibinfo {author} {\bibfnamefont {J{\"o}rg~B}\
  \bibnamefont {G{\"o}tte}}, \bibinfo {author} {\bibfnamefont {Stephen~M}\
  \bibnamefont {Barnett}}, \ and\ \bibinfo {author} {\bibfnamefont {Miles}\
  \bibnamefont {Padgett}},\ }\bibfield  {title} {\enquote {\bibinfo {title} {On
  the dragging of light by a rotating medium},}\ }\href@noop {} {\bibfield
  {journal} {\bibinfo  {journal} {Proceedings of the Royal Society A:
  Mathematical, Physical and Engineering Sciences}\ }\textbf {\bibinfo {volume}
  {463}},\ \bibinfo {pages} {2185--2194} (\bibinfo {year} {2007})}\BibitemShut
  {NoStop}%
\bibitem [{\citenamefont {Carusotto}\ \emph {et~al.}(2003)\citenamefont
  {Carusotto}, \citenamefont {Artoni}, \citenamefont {La~Rocca},\ and\
  \citenamefont {Bassani}}]{PhysRevA.68.063819}%
  \BibitemOpen
  \bibfield  {author} {\bibinfo {author} {\bibfnamefont {I.}~\bibnamefont
  {Carusotto}}, \bibinfo {author} {\bibfnamefont {M.}~\bibnamefont {Artoni}},
  \bibinfo {author} {\bibfnamefont {G.~C.}\ \bibnamefont {La~Rocca}}, \ and\
  \bibinfo {author} {\bibfnamefont {F.}~\bibnamefont {Bassani}},\ }\bibfield
  {title} {\enquote {\bibinfo {title} {Transverse fresnel-fizeau drag effects
  in strongly dispersive media},}\ }\href@noop {} {\bibfield  {journal}
  {\bibinfo  {journal} {Phys. Rev. A}\ }\textbf {\bibinfo {volume} {68}},\
  \bibinfo {pages} {063819} (\bibinfo {year} {2003})}\BibitemShut {NoStop}%
\bibitem [{\citenamefont {Leonhardt}\ and\ \citenamefont
  {Piwnicki}(2000)}]{leonhardt2000relativistic}%
  \BibitemOpen
  \bibfield  {author} {\bibinfo {author} {\bibfnamefont {Ulf}\ \bibnamefont
  {Leonhardt}}\ and\ \bibinfo {author} {\bibfnamefont {Paul}\ \bibnamefont
  {Piwnicki}},\ }\bibfield  {title} {\enquote {\bibinfo {title} {Relativistic
  effects of light in moving media with extremely low group velocity},}\
  }\href@noop {} {\bibfield  {journal} {\bibinfo  {journal} {Physical review
  letters}\ }\textbf {\bibinfo {volume} {84}},\ \bibinfo {pages} {822}
  (\bibinfo {year} {2000})}\BibitemShut {NoStop}%
\bibitem [{\citenamefont {Khan}\ \emph {et~al.}(2021)\citenamefont {Khan},
  \citenamefont {Ullah}, \citenamefont {Jabar},\ and\ \citenamefont
  {Bacha}}]{khan2021fizeau}%
  \BibitemOpen
  \bibfield  {author} {\bibinfo {author} {\bibfnamefont {Ayub}\ \bibnamefont
  {Khan}}, \bibinfo {author} {\bibfnamefont {Sayed~Arif}\ \bibnamefont
  {Ullah}}, \bibinfo {author} {\bibfnamefont {MS~Abdul}\ \bibnamefont {Jabar}},
  \ and\ \bibinfo {author} {\bibfnamefont {Bakht~Amin}\ \bibnamefont {Bacha}},\
  }\bibfield  {title} {\enquote {\bibinfo {title} {Fizeau's light birefringence
  dragging effect in a moving chiral medium},}\ }\href@noop {} {\bibfield
  {journal} {\bibinfo  {journal} {The European Physical Journal Plus}\ }\textbf
  {\bibinfo {volume} {136}},\ \bibinfo {pages} {1--13} (\bibinfo {year}
  {2021})}\BibitemShut {NoStop}%
\bibitem [{\citenamefont {Qin}\ \emph {et~al.}(2020)\citenamefont {Qin},
  \citenamefont {Yang}, \citenamefont {Zhang}, \citenamefont {Chen},
  \citenamefont {Shen}, \citenamefont {Liu}, \citenamefont {Chen},
  \citenamefont {Jiang}, \citenamefont {Chen},\ and\ \citenamefont
  {Wan}}]{qin2020fast}%
  \BibitemOpen
  \bibfield  {author} {\bibinfo {author} {\bibfnamefont {Tian}\ \bibnamefont
  {Qin}}, \bibinfo {author} {\bibfnamefont {Jianfan}\ \bibnamefont {Yang}},
  \bibinfo {author} {\bibfnamefont {Fangxing}\ \bibnamefont {Zhang}}, \bibinfo
  {author} {\bibfnamefont {Yao}\ \bibnamefont {Chen}}, \bibinfo {author}
  {\bibfnamefont {Dongyi}\ \bibnamefont {Shen}}, \bibinfo {author}
  {\bibfnamefont {Wei}\ \bibnamefont {Liu}}, \bibinfo {author} {\bibfnamefont
  {Lei}\ \bibnamefont {Chen}}, \bibinfo {author} {\bibfnamefont {Xiaoshun}\
  \bibnamefont {Jiang}}, \bibinfo {author} {\bibfnamefont {Xianfeng}\
  \bibnamefont {Chen}}, \ and\ \bibinfo {author} {\bibfnamefont {Wenjie}\
  \bibnamefont {Wan}},\ }\bibfield  {title} {\enquote {\bibinfo {title}
  {Fast-and slow-light-enhanced light drag in a moving microcavity},}\
  }\href@noop {} {\bibfield  {journal} {\bibinfo  {journal} {Communications
  Physics}\ }\textbf {\bibinfo {volume} {3}},\ \bibinfo {pages} {1--8}
  (\bibinfo {year} {2020})}\BibitemShut {NoStop}%
\bibitem [{\citenamefont {Chiao}\ \emph {et~al.}(1964)\citenamefont {Chiao},
  \citenamefont {Garmire},\ and\ \citenamefont {Townes}}]{chiao1964self}%
  \BibitemOpen
  \bibfield  {author} {\bibinfo {author} {\bibfnamefont {Raymond~Y}\
  \bibnamefont {Chiao}}, \bibinfo {author} {\bibfnamefont {E}~\bibnamefont
  {Garmire}}, \ and\ \bibinfo {author} {\bibfnamefont {Charles~H}\ \bibnamefont
  {Townes}},\ }\bibfield  {title} {\enquote {\bibinfo {title} {Self-trapping of
  optical beams},}\ }\href@noop {} {\bibfield  {journal} {\bibinfo  {journal}
  {Physical Review Letters}\ }\textbf {\bibinfo {volume} {13}},\ \bibinfo
  {pages} {479} (\bibinfo {year} {1964})}\BibitemShut {NoStop}%
\bibitem [{\citenamefont {Fibich}\ and\ \citenamefont
  {Gaeta}(2000)}]{fibich2000critical}%
  \BibitemOpen
  \bibfield  {author} {\bibinfo {author} {\bibfnamefont {Gadi}\ \bibnamefont
  {Fibich}}\ and\ \bibinfo {author} {\bibfnamefont {Alexander~L}\ \bibnamefont
  {Gaeta}},\ }\bibfield  {title} {\enquote {\bibinfo {title} {Critical power
  for self-focusing in bulk media and in hollow waveguides},}\ }\href@noop {}
  {\bibfield  {journal} {\bibinfo  {journal} {Optics letters}\ }\textbf
  {\bibinfo {volume} {25}},\ \bibinfo {pages} {335--337} (\bibinfo {year}
  {2000})}\BibitemShut {NoStop}%
\bibitem [{\citenamefont {Meyer}\ and\ \citenamefont
  {Amer}(1990)}]{meyer1990optical}%
  \BibitemOpen
  \bibfield  {author} {\bibinfo {author} {\bibfnamefont {Gerhard}\ \bibnamefont
  {Meyer}}\ and\ \bibinfo {author} {\bibfnamefont {Nabil~M}\ \bibnamefont
  {Amer}},\ }\bibfield  {title} {\enquote {\bibinfo {title}
  {Optical-beam-deflection atomic force microscopy: The nacl (001) surface},}\
  }\href@noop {} {\bibfield  {journal} {\bibinfo  {journal} {Applied Physics
  Letters}\ }\textbf {\bibinfo {volume} {56}},\ \bibinfo {pages} {2100--2101}
  (\bibinfo {year} {1990})}\BibitemShut {NoStop}%
\bibitem [{\citenamefont {Dixon}\ \emph {et~al.}(2009)\citenamefont {Dixon},
  \citenamefont {Starling}, \citenamefont {Jordan},\ and\ \citenamefont
  {Howell}}]{dixon2009ultrasensitive}%
  \BibitemOpen
  \bibfield  {author} {\bibinfo {author} {\bibfnamefont {P~Ben}\ \bibnamefont
  {Dixon}}, \bibinfo {author} {\bibfnamefont {David~J}\ \bibnamefont
  {Starling}}, \bibinfo {author} {\bibfnamefont {Andrew~N}\ \bibnamefont
  {Jordan}}, \ and\ \bibinfo {author} {\bibfnamefont {John~C}\ \bibnamefont
  {Howell}},\ }\bibfield  {title} {\enquote {\bibinfo {title} {Ultrasensitive
  beam deflection measurement via interferometric weak value amplification},}\
  }\href@noop {} {\bibfield  {journal} {\bibinfo  {journal} {Physical review
  letters}\ }\textbf {\bibinfo {volume} {102}},\ \bibinfo {pages} {173601}
  (\bibinfo {year} {2009})}\BibitemShut {NoStop}%
\bibitem [{\citenamefont {Gibson}\ \emph {et~al.}(1970)\citenamefont {Gibson},
  \citenamefont {Kimmitt},\ and\ \citenamefont {Walker}}]{gibson1970photon}%
  \BibitemOpen
  \bibfield  {author} {\bibinfo {author} {\bibfnamefont {AF}~\bibnamefont
  {Gibson}}, \bibinfo {author} {\bibfnamefont {MF}~\bibnamefont {Kimmitt}}, \
  and\ \bibinfo {author} {\bibfnamefont {AC}~\bibnamefont {Walker}},\
  }\bibfield  {title} {\enquote {\bibinfo {title} {Photon drag in germanium},}\
  }\href@noop {} {\bibfield  {journal} {\bibinfo  {journal} {Applied Physics
  Letters}\ }\textbf {\bibinfo {volume} {17}},\ \bibinfo {pages} {75--77}
  (\bibinfo {year} {1970})}\BibitemShut {NoStop}%
\bibitem [{\citenamefont {Yee}(1972)}]{yee1972theory}%
  \BibitemOpen
  \bibfield  {author} {\bibinfo {author} {\bibfnamefont {Jick~H}\ \bibnamefont
  {Yee}},\ }\bibfield  {title} {\enquote {\bibinfo {title} {Theory of
  photon-drag effect in polar crystals},}\ }\href@noop {} {\bibfield  {journal}
  {\bibinfo  {journal} {Physical Review B}\ }\textbf {\bibinfo {volume} {6}},\
  \bibinfo {pages} {2279} (\bibinfo {year} {1972})}\BibitemShut {NoStop}%
\bibitem [{\citenamefont {Gibson}\ and\ \citenamefont
  {Kimmitt}(1980)}]{gibson1980photon}%
  \BibitemOpen
  \bibfield  {author} {\bibinfo {author} {\bibfnamefont {AF}~\bibnamefont
  {Gibson}}\ and\ \bibinfo {author} {\bibfnamefont {MF}~\bibnamefont
  {Kimmitt}},\ }\bibfield  {title} {\enquote {\bibinfo {title} {Photon drag
  detection},}\ }\href@noop {} {\bibfield  {journal} {\bibinfo  {journal}
  {Infrared and Millimeter Waves}\ }\textbf {\bibinfo {volume} {3}},\ \bibinfo
  {pages} {181--217} (\bibinfo {year} {1980})}\BibitemShut {NoStop}%
\bibitem [{\citenamefont {Grinberg}\ and\ \citenamefont
  {Luryi}(1988)}]{grinberg1988theory}%
  \BibitemOpen
  \bibfield  {author} {\bibinfo {author} {\bibfnamefont {Anatoly~A}\
  \bibnamefont {Grinberg}}\ and\ \bibinfo {author} {\bibfnamefont {Serge}\
  \bibnamefont {Luryi}},\ }\bibfield  {title} {\enquote {\bibinfo {title}
  {Theory of the photon-drag effect in a two-dimensional electron gas},}\
  }\href@noop {} {\bibfield  {journal} {\bibinfo  {journal} {Physical Review
  B}\ }\textbf {\bibinfo {volume} {38}},\ \bibinfo {pages} {87} (\bibinfo
  {year} {1988})}\BibitemShut {NoStop}%
\bibitem [{\citenamefont {Wieck}\ \emph {et~al.}(1990)\citenamefont {Wieck},
  \citenamefont {Sigg},\ and\ \citenamefont {Ploog}}]{wieck1990observation}%
  \BibitemOpen
  \bibfield  {author} {\bibinfo {author} {\bibfnamefont {AD}~\bibnamefont
  {Wieck}}, \bibinfo {author} {\bibfnamefont {H}~\bibnamefont {Sigg}}, \ and\
  \bibinfo {author} {\bibfnamefont {K}~\bibnamefont {Ploog}},\ }\bibfield
  {title} {\enquote {\bibinfo {title} {Observation of resonant photon drag in a
  two-dimensional electron gas},}\ }\href@noop {} {\bibfield  {journal}
  {\bibinfo  {journal} {Physical review letters}\ }\textbf {\bibinfo {volume}
  {64}},\ \bibinfo {pages} {463} (\bibinfo {year} {1990})}\BibitemShut
  {NoStop}%
\bibitem [{\citenamefont {Strekalov}\ \emph {et~al.}(2004)\citenamefont
  {Strekalov}, \citenamefont {Matsko}, \citenamefont {Yu},\ and\ \citenamefont
  {Maleki}}]{strekalov2004observation}%
  \BibitemOpen
  \bibfield  {author} {\bibinfo {author} {\bibfnamefont {Dmitry}\ \bibnamefont
  {Strekalov}}, \bibinfo {author} {\bibfnamefont {Andrey~B}\ \bibnamefont
  {Matsko}}, \bibinfo {author} {\bibfnamefont {Nan}\ \bibnamefont {Yu}}, \ and\
  \bibinfo {author} {\bibfnamefont {Lute}\ \bibnamefont {Maleki}},\ }\bibfield
  {title} {\enquote {\bibinfo {title} {Observation of light dragging in a
  rubidium vapor cell},}\ }\href@noop {} {\bibfield  {journal} {\bibinfo
  {journal} {Physical review letters}\ }\textbf {\bibinfo {volume} {93}},\
  \bibinfo {pages} {023601} (\bibinfo {year} {2004})}\BibitemShut {NoStop}%
\bibitem [{\citenamefont {Davuluri}\ and\ \citenamefont
  {Rostovtsev}(2012)}]{davuluri2012controllable}%
  \BibitemOpen
  \bibfield  {author} {\bibinfo {author} {\bibfnamefont {Sankar}\ \bibnamefont
  {Davuluri}}\ and\ \bibinfo {author} {\bibfnamefont {Yuri~V}\ \bibnamefont
  {Rostovtsev}},\ }\bibfield  {title} {\enquote {\bibinfo {title} {Controllable
  enhanced dragging of light in ultradispersive media},}\ }\href@noop {}
  {\bibfield  {journal} {\bibinfo  {journal} {Physical Review A}\ }\textbf
  {\bibinfo {volume} {86}},\ \bibinfo {pages} {013806} (\bibinfo {year}
  {2012})}\BibitemShut {NoStop}%
\bibitem [{\citenamefont {Kazemi}\ and\ \citenamefont
  {Mahmoudi}(2018)}]{kazemi2018phase}%
  \BibitemOpen
  \bibfield  {author} {\bibinfo {author} {\bibfnamefont {Seyedeh~Hamideh}\
  \bibnamefont {Kazemi}}\ and\ \bibinfo {author} {\bibfnamefont {Mohammad}\
  \bibnamefont {Mahmoudi}},\ }\bibfield  {title} {\enquote {\bibinfo {title}
  {Phase-controlled photon drag in a slow-light moving medium},}\ }\href@noop
  {} {\bibfield  {journal} {\bibinfo  {journal} {arXiv preprint
  arXiv:1810.00560}\ } (\bibinfo {year} {2018})}\BibitemShut {NoStop}%
\bibitem [{\citenamefont {Franke-Arnold}\ \emph {et~al.}(2011)\citenamefont
  {Franke-Arnold}, \citenamefont {Gibson}, \citenamefont {Boyd},\ and\
  \citenamefont {Padgett}}]{franke2011rotary}%
  \BibitemOpen
  \bibfield  {author} {\bibinfo {author} {\bibfnamefont {Sonja}\ \bibnamefont
  {Franke-Arnold}}, \bibinfo {author} {\bibfnamefont {Graham}\ \bibnamefont
  {Gibson}}, \bibinfo {author} {\bibfnamefont {Robert~W}\ \bibnamefont {Boyd}},
  \ and\ \bibinfo {author} {\bibfnamefont {Miles~J}\ \bibnamefont {Padgett}},\
  }\bibfield  {title} {\enquote {\bibinfo {title} {Rotary photon drag enhanced
  by a slow-light medium},}\ }\href@noop {} {\bibfield  {journal} {\bibinfo
  {journal} {Science}\ }\textbf {\bibinfo {volume} {333}},\ \bibinfo {pages}
  {65--67} (\bibinfo {year} {2011})}\BibitemShut {NoStop}%
\bibitem [{\citenamefont {Safari}\ \emph {et~al.}(2016)\citenamefont {Safari},
  \citenamefont {De~Leon}, \citenamefont {Mirhosseini}, \citenamefont
  {Maga{\~n}a-Loaiza},\ and\ \citenamefont {Boyd}}]{safari2016light}%
  \BibitemOpen
  \bibfield  {author} {\bibinfo {author} {\bibfnamefont {Akbar}\ \bibnamefont
  {Safari}}, \bibinfo {author} {\bibfnamefont {Israel}\ \bibnamefont
  {De~Leon}}, \bibinfo {author} {\bibfnamefont {Mohammad}\ \bibnamefont
  {Mirhosseini}}, \bibinfo {author} {\bibfnamefont {Omar~S}\ \bibnamefont
  {Maga{\~n}a-Loaiza}}, \ and\ \bibinfo {author} {\bibfnamefont {Robert~W}\
  \bibnamefont {Boyd}},\ }\bibfield  {title} {\enquote {\bibinfo {title}
  {Light-drag enhancement by a highly dispersive rubidium vapor},}\ }\href@noop
  {} {\bibfield  {journal} {\bibinfo  {journal} {Physical review letters}\
  }\textbf {\bibinfo {volume} {116}},\ \bibinfo {pages} {013601} (\bibinfo
  {year} {2016})}\BibitemShut {NoStop}%
\bibitem [{\citenamefont {Toll}(1956)}]{toll1956causality}%
  \BibitemOpen
  \bibfield  {author} {\bibinfo {author} {\bibfnamefont {John~S}\ \bibnamefont
  {Toll}},\ }\bibfield  {title} {\enquote {\bibinfo {title} {Causality and the
  dispersion relation: logical foundations},}\ }\href@noop {} {\bibfield
  {journal} {\bibinfo  {journal} {Physical review}\ }\textbf {\bibinfo {volume}
  {104}},\ \bibinfo {pages} {1760} (\bibinfo {year} {1956})}\BibitemShut
  {NoStop}%
\bibitem [{\citenamefont {Jones}(1972)}]{jones1972fresnel}%
  \BibitemOpen
  \bibfield  {author} {\bibinfo {author} {\bibfnamefont {Reginald~Victor}\
  \bibnamefont {Jones}},\ }\bibfield  {title} {\enquote {\bibinfo {title}
  {'fresnel aether drag'in a transversely moving medium},}\ }\href@noop {}
  {\bibfield  {journal} {\bibinfo  {journal} {Proceedings of the Royal Society
  of London. A. Mathematical and Physical Sciences}\ }\textbf {\bibinfo
  {volume} {328}},\ \bibinfo {pages} {337--352} (\bibinfo {year}
  {1972})}\BibitemShut {NoStop}%
\bibitem [{\citenamefont {Leach}\ \emph {et~al.}(2008)\citenamefont {Leach},
  \citenamefont {Wright}, \citenamefont {G{\"o}tte}, \citenamefont {Girkin},
  \citenamefont {Allen}, \citenamefont {Franke-Arnold}, \citenamefont
  {Barnett},\ and\ \citenamefont {Padgett}}]{leach2008aether}%
  \BibitemOpen
  \bibfield  {author} {\bibinfo {author} {\bibfnamefont {J}~\bibnamefont
  {Leach}}, \bibinfo {author} {\bibfnamefont {AJ}~\bibnamefont {Wright}},
  \bibinfo {author} {\bibfnamefont {JB}~\bibnamefont {G{\"o}tte}}, \bibinfo
  {author} {\bibfnamefont {JM}~\bibnamefont {Girkin}}, \bibinfo {author}
  {\bibfnamefont {L}~\bibnamefont {Allen}}, \bibinfo {author} {\bibfnamefont
  {S}~\bibnamefont {Franke-Arnold}}, \bibinfo {author} {\bibfnamefont
  {SM}~\bibnamefont {Barnett}}, \ and\ \bibinfo {author} {\bibfnamefont
  {MJ}~\bibnamefont {Padgett}},\ }\bibfield  {title} {\enquote {\bibinfo
  {title} {“aether drag” and moving images},}\ }\href@noop {} {\bibfield
  {journal} {\bibinfo  {journal} {Physical review letters}\ }\textbf {\bibinfo
  {volume} {100}},\ \bibinfo {pages} {153902} (\bibinfo {year}
  {2008})}\BibitemShut {NoStop}%
\bibitem [{\citenamefont {Bigelow}\ \emph {et~al.}(2003)\citenamefont
  {Bigelow}, \citenamefont {Lepeshkin},\ and\ \citenamefont
  {Boyd}}]{bigelow2003superluminal}%
  \BibitemOpen
  \bibfield  {author} {\bibinfo {author} {\bibfnamefont {Matthew~S}\
  \bibnamefont {Bigelow}}, \bibinfo {author} {\bibfnamefont {Nick~N}\
  \bibnamefont {Lepeshkin}}, \ and\ \bibinfo {author} {\bibfnamefont
  {Robert~W}\ \bibnamefont {Boyd}},\ }\bibfield  {title} {\enquote {\bibinfo
  {title} {Superluminal and slow light propagation in a room-temperature
  solid},}\ }\href@noop {} {\bibfield  {journal} {\bibinfo  {journal}
  {Science}\ }\textbf {\bibinfo {volume} {301}},\ \bibinfo {pages} {200--202}
  (\bibinfo {year} {2003})}\BibitemShut {NoStop}%
\bibitem [{\citenamefont {Safari}(2022)}]{safari2022group}%
  \BibitemOpen
  \bibfield  {author} {\bibinfo {author} {\bibfnamefont {Akbar}\ \bibnamefont
  {Safari}},\ }\bibfield  {title} {\enquote {\bibinfo {title} {manuscript in
  preparation},}\ }\href@noop {} {\bibfield  {journal} {\bibinfo  {journal}
  {TBD}\ } (\bibinfo {year} {2022})}\BibitemShut {NoStop}%
\bibitem [{\citenamefont {Boccia}\ \emph {et~al.}(2009)\citenamefont {Boccia},
  \citenamefont {Russo}, \citenamefont {Amendola},\ and\ \citenamefont
  {Di~Massa}}]{boccia2009tunable}%
  \BibitemOpen
  \bibfield  {author} {\bibinfo {author} {\bibfnamefont {L}~\bibnamefont
  {Boccia}}, \bibinfo {author} {\bibfnamefont {I}~\bibnamefont {Russo}},
  \bibinfo {author} {\bibfnamefont {G}~\bibnamefont {Amendola}}, \ and\
  \bibinfo {author} {\bibfnamefont {G}~\bibnamefont {Di~Massa}},\ }\bibfield
  {title} {\enquote {\bibinfo {title} {Tunable frequency-selective surfaces for
  beam-steering applications},}\ }\href@noop {} {\bibfield  {journal} {\bibinfo
   {journal} {Electronics letters}\ }\textbf {\bibinfo {volume} {45}},\
  \bibinfo {pages} {1213--1215} (\bibinfo {year} {2009})}\BibitemShut {NoStop}%
\bibitem [{\citenamefont {Golub}(1992)}]{golub1992beam}%
  \BibitemOpen
  \bibfield  {author} {\bibinfo {author} {\bibfnamefont {I}~\bibnamefont
  {Golub}},\ }\bibfield  {title} {\enquote {\bibinfo {title} {Beam deflection
  and ultrafast angular scanning by a time-varying optically induced prism},}\
  }\href@noop {} {\bibfield  {journal} {\bibinfo  {journal} {Optics
  communications}\ }\textbf {\bibinfo {volume} {94}},\ \bibinfo {pages}
  {143--146} (\bibinfo {year} {1992})}\BibitemShut {NoStop}%
\bibitem [{\citenamefont {Allen}\ \emph {et~al.}(1963)\citenamefont {Allen},
  \citenamefont {Brault},\ and\ \citenamefont {Moore}}]{allen1963new}%
  \BibitemOpen
  \bibfield  {author} {\bibinfo {author} {\bibfnamefont {Robert~D}\
  \bibnamefont {Allen}}, \bibinfo {author} {\bibfnamefont {James}\ \bibnamefont
  {Brault}}, \ and\ \bibinfo {author} {\bibfnamefont {Robert~D}\ \bibnamefont
  {Moore}},\ }\bibfield  {title} {\enquote {\bibinfo {title} {A new method of
  polarization microscopic analysis: I. scanning with a birefringence detection
  system},}\ }\href@noop {} {\bibfield  {journal} {\bibinfo  {journal} {The
  Journal of cell biology}\ }\textbf {\bibinfo {volume} {18}},\ \bibinfo
  {pages} {223--235} (\bibinfo {year} {1963})}\BibitemShut {NoStop}%
\bibitem [{\citenamefont {Gong}\ \emph {et~al.}(2010)\citenamefont {Gong},
  \citenamefont {Zhan}, \citenamefont {Liu} \emph {et~al.}}]{gong2010review}%
  \BibitemOpen
  \bibfield  {author} {\bibinfo {author} {\bibfnamefont {Jie-Qiong}\
  \bibnamefont {Gong}}, \bibinfo {author} {\bibfnamefont {Hai-Gang}\
  \bibnamefont {Zhan}}, \bibinfo {author} {\bibfnamefont {Da-Zhao}\
  \bibnamefont {Liu}},  \emph {et~al.},\ }\bibfield  {title} {\enquote
  {\bibinfo {title} {A review on polarization information in the remote sensing
  detection},}\ }\href@noop {} {\bibfield  {journal} {\bibinfo  {journal}
  {Spectroscopy and Spectral Analysis}\ }\textbf {\bibinfo {volume} {30}},\
  \bibinfo {pages} {1088--1095} (\bibinfo {year} {2010})}\BibitemShut {NoStop}%
\bibitem [{\citenamefont {Hasegawa}(2000)}]{hasegawa2000historical}%
  \BibitemOpen
  \bibfield  {author} {\bibinfo {author} {\bibfnamefont {Akira}\ \bibnamefont
  {Hasegawa}},\ }\bibfield  {title} {\enquote {\bibinfo {title} {An historical
  review of application of optical solitons for high speed communications},}\
  }\href@noop {} {\bibfield  {journal} {\bibinfo  {journal} {Chaos: An
  Interdisciplinary Journal of Nonlinear Science}\ }\textbf {\bibinfo {volume}
  {10}},\ \bibinfo {pages} {475--485} (\bibinfo {year} {2000})}\BibitemShut
  {NoStop}%
\bibitem [{\citenamefont {Ablowitz}\ \emph {et~al.}(2000)\citenamefont
  {Ablowitz}, \citenamefont {Biondini},\ and\ \citenamefont
  {Ostrovsky}}]{ablowitz2000optical}%
  \BibitemOpen
  \bibfield  {author} {\bibinfo {author} {\bibfnamefont {Mark~J}\ \bibnamefont
  {Ablowitz}}, \bibinfo {author} {\bibfnamefont {Gino}\ \bibnamefont
  {Biondini}}, \ and\ \bibinfo {author} {\bibfnamefont {Lev~A}\ \bibnamefont
  {Ostrovsky}},\ }\bibfield  {title} {\enquote {\bibinfo {title} {Optical
  solitons: perspectives and applications},}\ }\href@noop {} {\bibfield
  {journal} {\bibinfo  {journal} {Chaos: An Interdisciplinary Journal of
  Nonlinear Science}\ }\textbf {\bibinfo {volume} {10}},\ \bibinfo {pages}
  {471--474} (\bibinfo {year} {2000})}\BibitemShut {NoStop}%
\bibitem [{\citenamefont {Kivshar}\ and\ \citenamefont
  {Luther-Davies}(1998)}]{kivshar1998dark}%
  \BibitemOpen
  \bibfield  {author} {\bibinfo {author} {\bibfnamefont {Yuri~S}\ \bibnamefont
  {Kivshar}}\ and\ \bibinfo {author} {\bibfnamefont {Barry}\ \bibnamefont
  {Luther-Davies}},\ }\bibfield  {title} {\enquote {\bibinfo {title} {Dark
  optical solitons: physics and applications},}\ }\href@noop {} {\bibfield
  {journal} {\bibinfo  {journal} {Physics reports}\ }\textbf {\bibinfo {volume}
  {298}},\ \bibinfo {pages} {81--197} (\bibinfo {year} {1998})}\BibitemShut
  {NoStop}%
\bibitem [{\citenamefont {De~Oliveira}\ \emph {et~al.}(1992)\citenamefont
  {De~Oliveira}, \citenamefont {de~Moura}, \citenamefont {Hickmann},\ and\
  \citenamefont {Gomes}}]{de1992self}%
  \BibitemOpen
  \bibfield  {author} {\bibinfo {author} {\bibfnamefont {JR}~\bibnamefont
  {De~Oliveira}}, \bibinfo {author} {\bibfnamefont {Marco~A}\ \bibnamefont
  {de~Moura}}, \bibinfo {author} {\bibfnamefont {J~Miguel}\ \bibnamefont
  {Hickmann}}, \ and\ \bibinfo {author} {\bibfnamefont {ASL}\ \bibnamefont
  {Gomes}},\ }\bibfield  {title} {\enquote {\bibinfo {title} {Self-steepening
  of optical pulses in dispersive media},}\ }\href@noop {} {\bibfield
  {journal} {\bibinfo  {journal} {JOSA B}\ }\textbf {\bibinfo {volume} {9}},\
  \bibinfo {pages} {2025--2027} (\bibinfo {year} {1992})}\BibitemShut {NoStop}%
\bibitem [{\citenamefont {Hern{\'a}ndez}\ \emph {et~al.}(2020)\citenamefont
  {Hern{\'a}ndez}, \citenamefont {Bonetti}, \citenamefont {Linale},
  \citenamefont {Grosz},\ and\ \citenamefont {Fierens}}]{hernandez2020soliton}%
  \BibitemOpen
  \bibfield  {author} {\bibinfo {author} {\bibfnamefont {Santiago~M}\
  \bibnamefont {Hern{\'a}ndez}}, \bibinfo {author} {\bibfnamefont
  {J}~\bibnamefont {Bonetti}}, \bibinfo {author} {\bibfnamefont
  {N}~\bibnamefont {Linale}}, \bibinfo {author} {\bibfnamefont
  {Diego~Fernando}\ \bibnamefont {Grosz}}, \ and\ \bibinfo {author}
  {\bibfnamefont {Pablo~Ignacio}\ \bibnamefont {Fierens}},\ }\bibfield  {title}
  {\enquote {\bibinfo {title} {Soliton solutions and self-steepening in the
  photon-conserving nonlinear schr{\"o}dinger equation},}\ }\href@noop {}
  {\bibfield  {journal} {\bibinfo  {journal} {Waves in Random and Complex
  Media}\ ,\ \bibinfo {pages} {1--17}} (\bibinfo {year} {2020})}\BibitemShut
  {NoStop}%
\bibitem [{\citenamefont {Marcucci}\ \emph {et~al.}(2019)\citenamefont
  {Marcucci}, \citenamefont {Pierangeli}, \citenamefont {Gentilini},
  \citenamefont {Ghofraniha}, \citenamefont {Chen},\ and\ \citenamefont
  {Conti}}]{marcucci2019optical}%
  \BibitemOpen
  \bibfield  {author} {\bibinfo {author} {\bibfnamefont {Giulia}\ \bibnamefont
  {Marcucci}}, \bibinfo {author} {\bibfnamefont {Davide}\ \bibnamefont
  {Pierangeli}}, \bibinfo {author} {\bibfnamefont {Silvia}\ \bibnamefont
  {Gentilini}}, \bibinfo {author} {\bibfnamefont {Neda}\ \bibnamefont
  {Ghofraniha}}, \bibinfo {author} {\bibfnamefont {Zhigang}\ \bibnamefont
  {Chen}}, \ and\ \bibinfo {author} {\bibfnamefont {Claudio}\ \bibnamefont
  {Conti}},\ }\bibfield  {title} {\enquote {\bibinfo {title} {Optical spatial
  shock waves in nonlocal nonlinear media},}\ }\href@noop {} {\bibfield
  {journal} {\bibinfo  {journal} {Advances in Physics: X}\ }\textbf {\bibinfo
  {volume} {4}},\ \bibinfo {pages} {1662733} (\bibinfo {year}
  {2019})}\BibitemShut {NoStop}%
\bibitem [{\citenamefont {Conti}\ \emph {et~al.}(2005)\citenamefont {Conti},
  \citenamefont {Peccianti},\ and\ \citenamefont {Assanto}}]{conti2005spatial}%
  \BibitemOpen
  \bibfield  {author} {\bibinfo {author} {\bibfnamefont {Claudio}\ \bibnamefont
  {Conti}}, \bibinfo {author} {\bibfnamefont {Marco}\ \bibnamefont
  {Peccianti}}, \ and\ \bibinfo {author} {\bibfnamefont {Gaetano}\ \bibnamefont
  {Assanto}},\ }\bibfield  {title} {\enquote {\bibinfo {title} {Spatial
  solitons and modulational instability in the presence of large birefringence:
  The case of highly nonlocal liquid crystals},}\ }\href@noop {} {\bibfield
  {journal} {\bibinfo  {journal} {Physical Review E}\ }\textbf {\bibinfo
  {volume} {72}},\ \bibinfo {pages} {066614} (\bibinfo {year}
  {2005})}\BibitemShut {NoStop}%
\bibitem [{\citenamefont {Doiron}\ and\ \citenamefont
  {Hach{\'e}}(2004)}]{doiron2004time}%
  \BibitemOpen
  \bibfield  {author} {\bibinfo {author} {\bibfnamefont {Serge}\ \bibnamefont
  {Doiron}}\ and\ \bibinfo {author} {\bibfnamefont {Alain}\ \bibnamefont
  {Hach{\'e}}},\ }\bibfield  {title} {\enquote {\bibinfo {title} {Time
  evolution of reflective thermal lenses and measurement of thermal diffusivity
  in bulk solids},}\ }\href@noop {} {\bibfield  {journal} {\bibinfo  {journal}
  {Applied optics}\ }\textbf {\bibinfo {volume} {43}},\ \bibinfo {pages}
  {4250--4253} (\bibinfo {year} {2004})}\BibitemShut {NoStop}%
\bibitem [{\citenamefont {Hogan}\ \emph {et~al.}(2022)\citenamefont {Hogan},
  \citenamefont {Marcucci}, \citenamefont {Safari}, \citenamefont {Black},
  \citenamefont {Braverman}, \citenamefont {Upham},\ and\ \citenamefont
  {Boyd}}]{hogan2022nlphoton}%
  \BibitemOpen
  \bibfield  {author} {\bibinfo {author} {\bibfnamefont {Ryan}\ \bibnamefont
  {Hogan}}, \bibinfo {author} {\bibfnamefont {G}~\bibnamefont {Marcucci}},
  \bibinfo {author} {\bibfnamefont {Akbar}\ \bibnamefont {Safari}}, \bibinfo
  {author} {\bibfnamefont {Nicholas~A.}\ \bibnamefont {Black}}, \bibinfo
  {author} {\bibfnamefont {Boris}\ \bibnamefont {Braverman}}, \bibinfo {author}
  {\bibfnamefont {Jeremy}\ \bibnamefont {Upham}}, \ and\ \bibinfo {author}
  {\bibfnamefont {Robert~W.}\ \bibnamefont {Boyd}},\ }\bibfield  {title}
  {\enquote {\bibinfo {title} {manuscript in preparation},}\ }\href@noop {}
  {\bibfield  {journal} {\bibinfo  {journal} {TBD}\ } (\bibinfo {year}
  {2022})}\BibitemShut {NoStop}%
\bibitem [{\citenamefont {Lee}\ and\ \citenamefont
  {Lee}(1990)}]{lee1990measurements}%
  \BibitemOpen
  \bibfield  {author} {\bibinfo {author} {\bibfnamefont {Hak~Kyu}\ \bibnamefont
  {Lee}}\ and\ \bibinfo {author} {\bibfnamefont {Sang~Soo}\ \bibnamefont
  {Lee}},\ }\bibfield  {title} {\enquote {\bibinfo {title} {Measurements of the
  anisotropic nonlinear refractive-index coefficients of ruby},}\ }\href@noop
  {} {\bibfield  {journal} {\bibinfo  {journal} {Optics letters}\ }\textbf
  {\bibinfo {volume} {15}},\ \bibinfo {pages} {54--56} (\bibinfo {year}
  {1990})}\BibitemShut {NoStop}%
\bibitem [{\citenamefont {Moll}\ \emph {et~al.}(2003)\citenamefont {Moll},
  \citenamefont {Gaeta},\ and\ \citenamefont {Fibich}}]{moll2003self}%
  \BibitemOpen
  \bibfield  {author} {\bibinfo {author} {\bibfnamefont {KD}~\bibnamefont
  {Moll}}, \bibinfo {author} {\bibfnamefont {Alexander~L}\ \bibnamefont
  {Gaeta}}, \ and\ \bibinfo {author} {\bibfnamefont {Gadi}\ \bibnamefont
  {Fibich}},\ }\bibfield  {title} {\enquote {\bibinfo {title} {Self-similar
  optical wave collapse: observation of the townes profile},}\ }\href@noop {}
  {\bibfield  {journal} {\bibinfo  {journal} {Physical review letters}\
  }\textbf {\bibinfo {volume} {90}},\ \bibinfo {pages} {203902} (\bibinfo
  {year} {2003})}\BibitemShut {NoStop}%
\end{thebibliography}%
\newpage
\clearpage
\section*{Supplementary Materials}

\subsection*{Transverse shift at different longitudinal positions}

\begin{figure}[H]                   %Fig2
\centering
\includegraphics[width=0.95\linewidth]{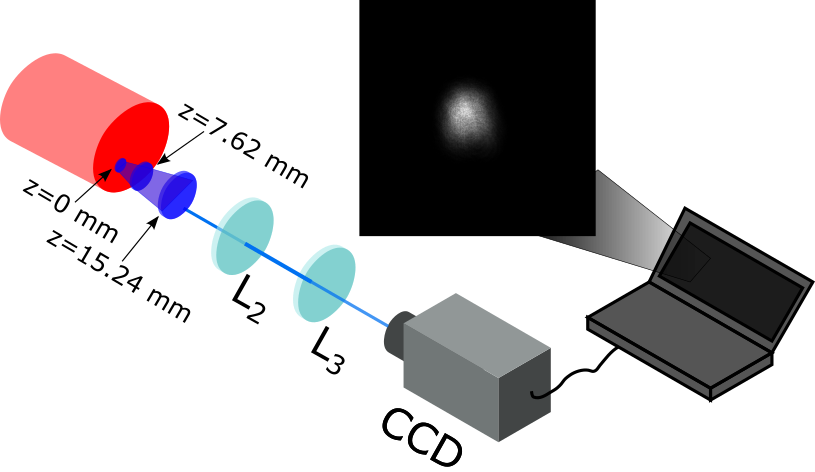} 
\caption{A schematic showing the three positions imaged by translating a CCD fast camera with a frame rate of 1000 frames/s to three positions moving away from the back face of the crystal by using a 4-f system of lens 2 and 3.  A single frame of a video of the beam at the output face (i.e. $z = 0$ mm) is shown in the inset of which the COM is taken to show the movement of the COM and the amount the beam is dragged over many frames. The frame shows a large beam that encompasses both the o- and e- beams as they are of the beam enlarges significantly propagating through 2-cm in the ruby crystal.}
\label{Figure1S} 
\end{figure}

Transverse drag, centre of mass (COM) trajectories, and the output angle were measured at the crystal back face ($z=0$ mm) for a variety of rotation speeds in the linear ($P_0 = 0.2$ mW), nonlinear ($P_0 = 100$ mW), and highly nonlinear regimes ($P_0 = 520$ mW). A schematic of these positions can be seen in Fig. \ref{Figure1S}. The CCD can image at different points using a translation stage with range of $\pm$ 25.4 mm from $z=0$. A set of measurements was taken at each z position, consisting of three powers, $P_0 = 0.2$ mW, $P_0 = 100$ mW, and $P_0 = 520$ mW for rotation speeds between 1-9000 deg/s. The data sets were then analyzed to extract the transverse shift, the output angle, and COM trajectories. The output angle was calculated by measuring the beam position at several z-planes and the amount of shift along y to find the angle. The angle is directly calculated 
\begin{equation}
    \theta_{L}=\arctan(\frac{\Delta y}{\Delta z}).
\end{equation}

\begin{figure}[t!]                 %Fig6
\centering
\includegraphics[width=0.95\linewidth]{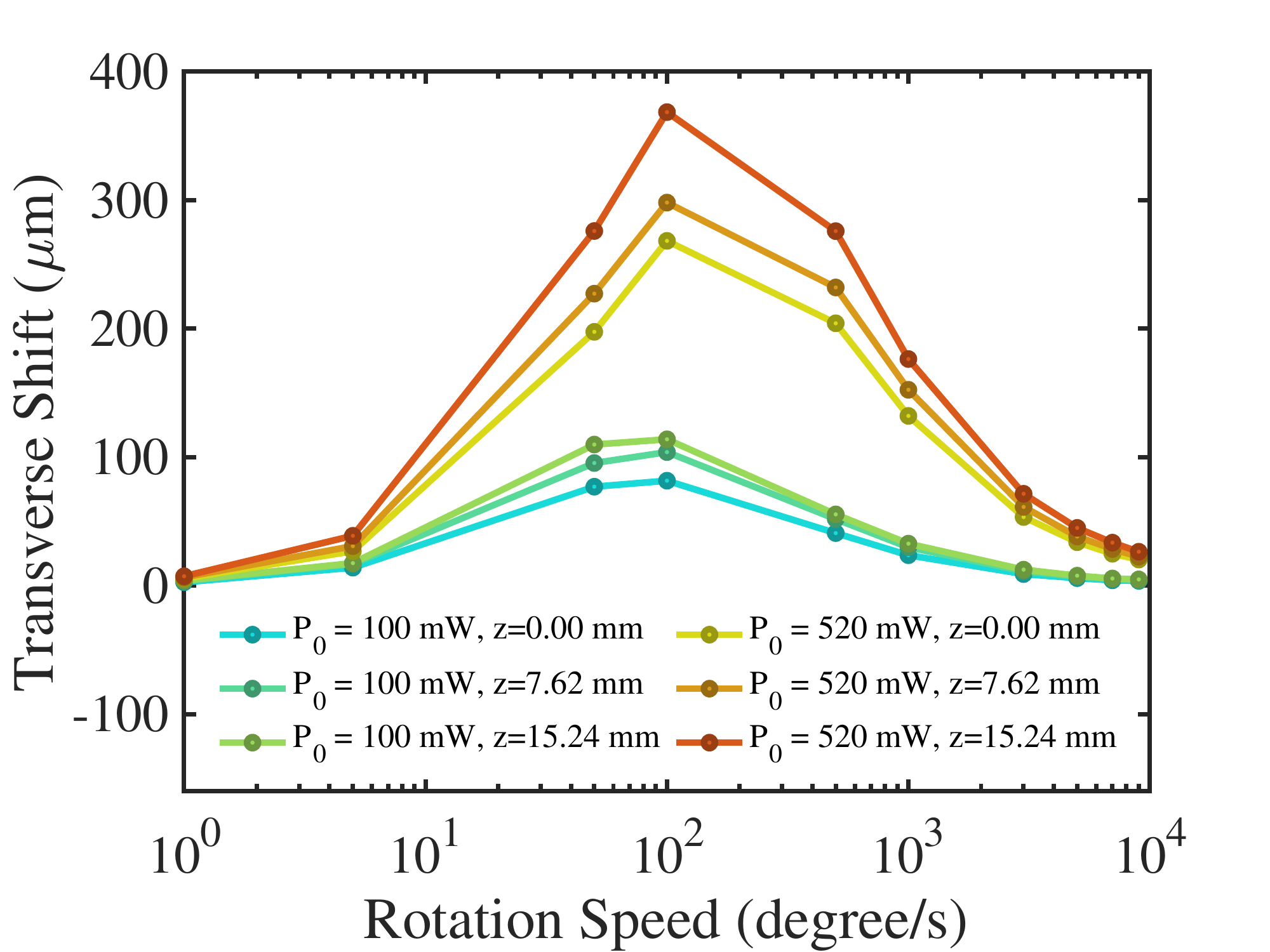} 
\caption{Experimentally measured transverse shift in nonlinear ($P_0 = 100$ mW), and highly nonlinear ($P_0 = 520$ mW) regimes at different z-positions. Measurements were taken at $z=0$, $z=7.62$ mm, and $15.24$ mm to calculate output angle. The transverse shift approaches $\Delta y = 10-15$ $\mu$m for an input power $P_0 = 100$ mW, and$\Delta y = 60$ $\mu$m for $P_0 = 520$ mW. The curve takes shape similar to a log-normal distribution, but modelled as the sum of two decaying exponentials with different decay rates centered around a rotation speed of $\Omega=100$ deg/s.} 
\label{Figure2S} 
\end{figure}

\begin{figure}[t!]
\centering
\includegraphics[width=0.95\linewidth]{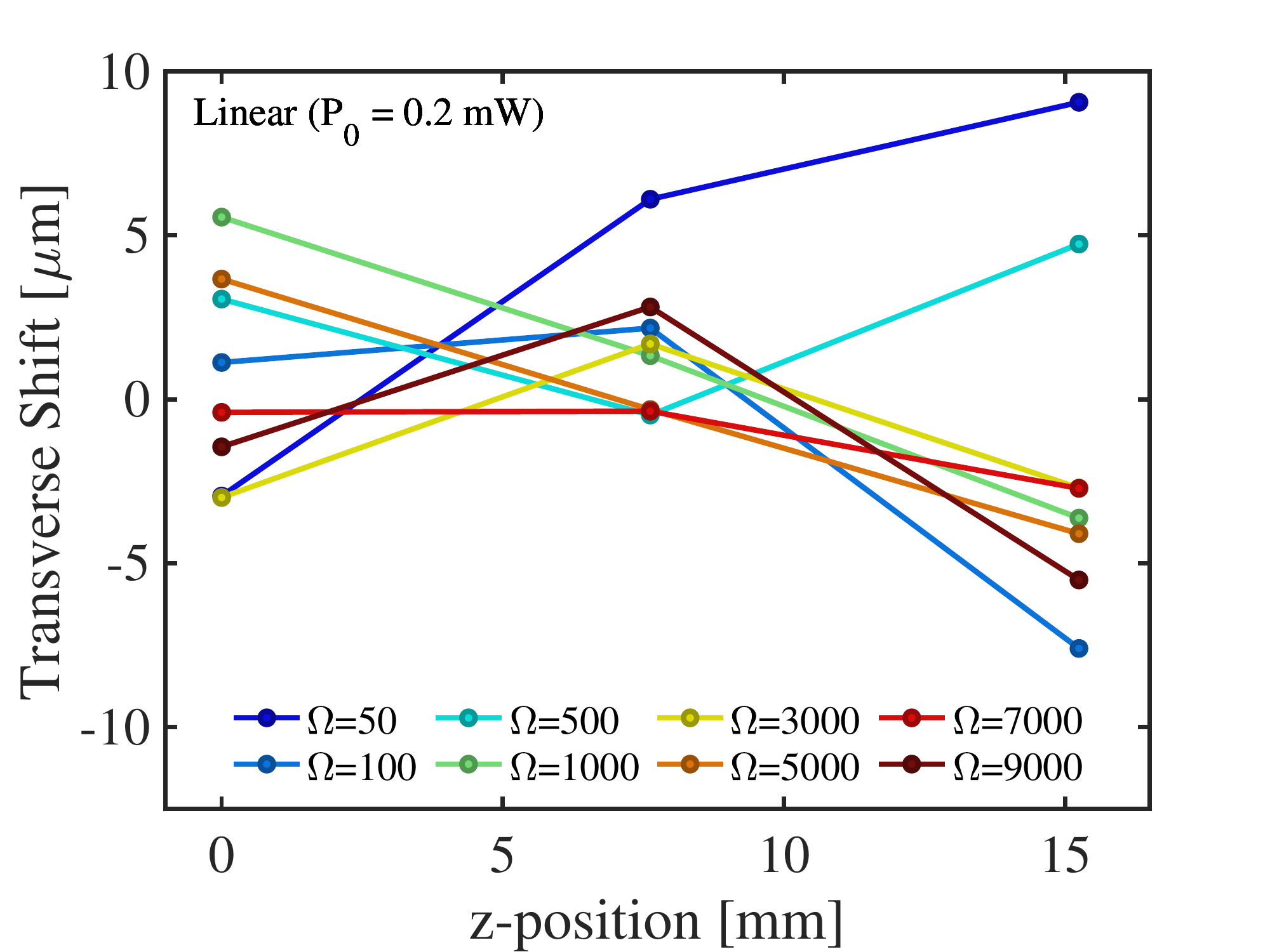}
\caption{Extrapolation of the amount of transverse shift of three experimentally measured points along the propagation direction is plotted for an input laser power of $P_0 = 0.2$ mW. Evolution of amount of transverse drag at three points including the crystal back face, and two positions hereafter, as shown in Figure \ref{Figure1S}. Values are enhanced by a factor of 10 for plotting purposes, where $\Delta y \rightarrow \Delta y\times10$. One can see that the relationship between the transverse shift and the propagation distance along z after the crystal in the linear regime does not show a distinct behaviour other than random movement due to the noise of the system. Indeed, the linear regime behaviour is akin to the noise. The deviation comes from a change in the output angle due to the nonlinear response of the crystal on the COM as it propagates through the crystal.
}
\label{Figure3aS}
\end{figure}

\begin{figure}[t!]
  \includegraphics[width=0.99\linewidth]{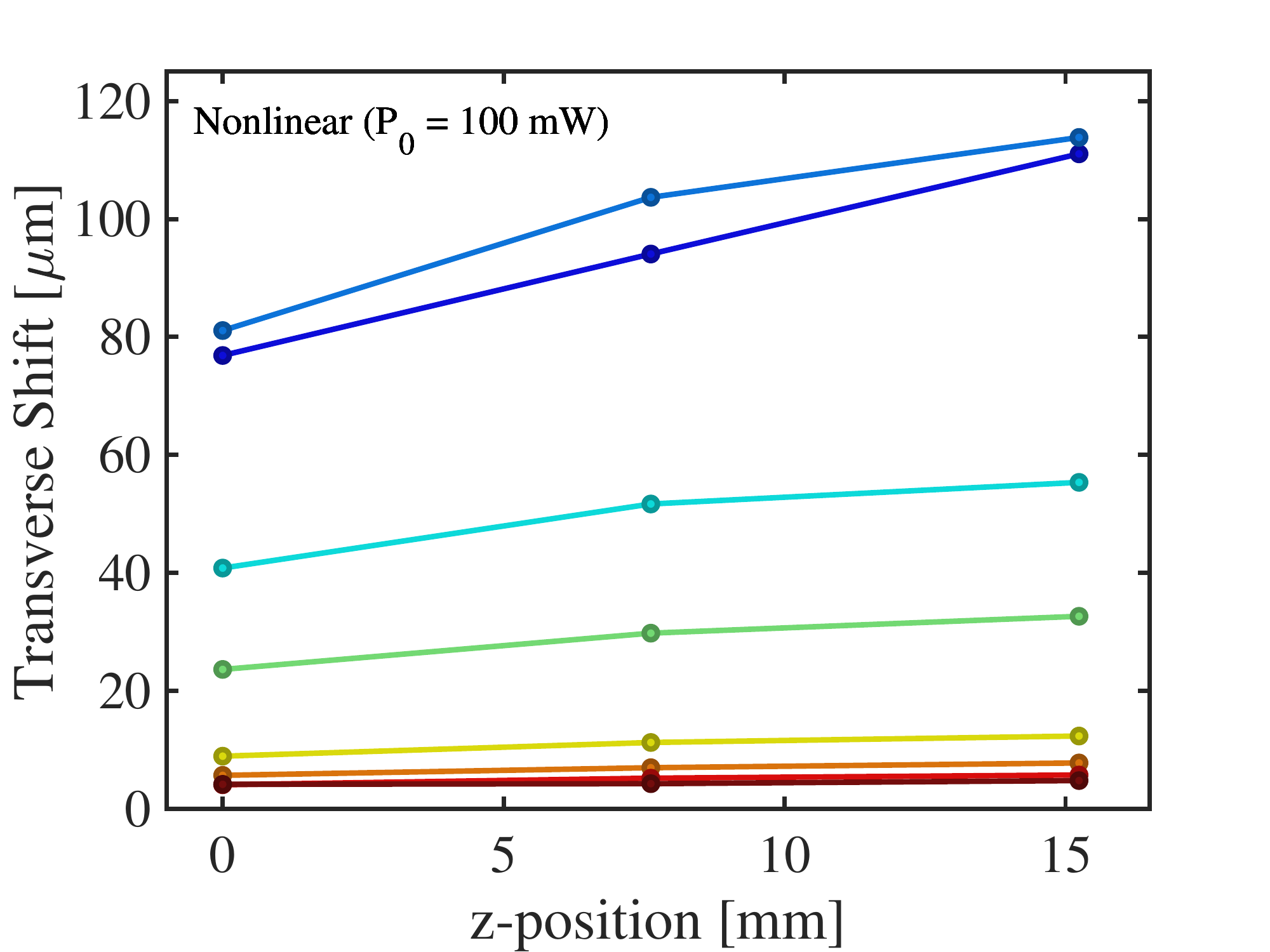}
  \includegraphics[width=0.99\linewidth]{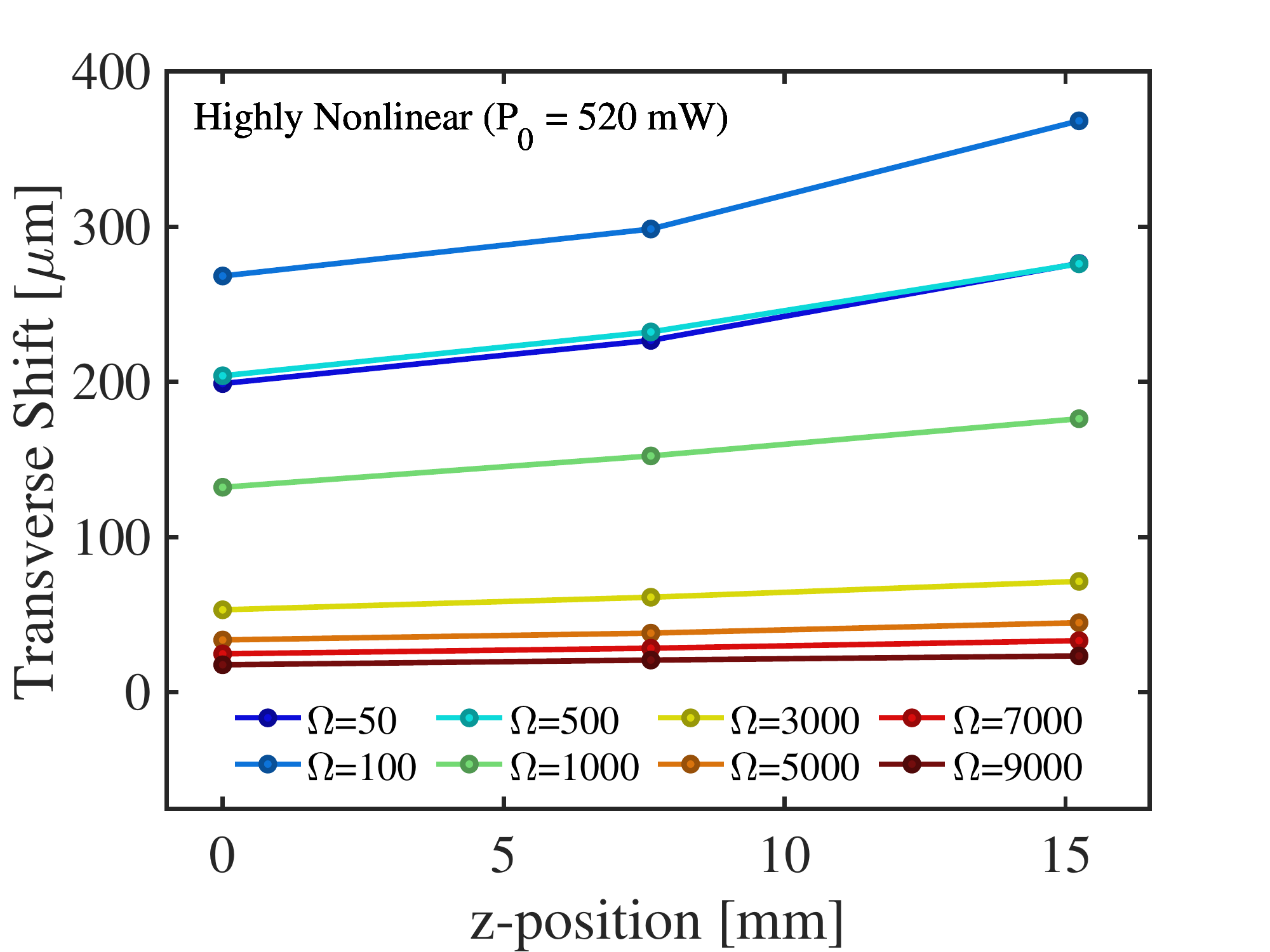} 
\caption{Extrapolation of transverse shift for input laser powers  of $P_0 = 100$ mW, and $P_0 = 520$ mW. Evolution of amount of transverse drag at three points including the crystal back face, and two positions hereafter, as shown in Fig. \ref{Figure1S}. The extrapolation of these points in the highly nonlinear regime also show a linear dependence on the transverse shift as the propagation distance increases, consistent with a straight-line propagation of the COM. The difference from the nonlinear regime is the magnitude of the slopes are much larger as a consequence of a larger nonlinear response in the system for input powers of 520 mW. One could extrapolate these curves as a linear regression back to the crystal front face $z=-20$ mm and see that the value does not reach zero. It is clear in the range of speeds from $\Omega=50 - 1000$ deg/s, where the value would be non-zero at the crystal front face, and thus a nonlinear trajectory is suspected.}
\label{Figure3S}
\end{figure}

Using the values for the transverse shifts as $\Delta y$ and the difference between the z-positions along propagation $\Delta z$, one can calculate the output angle seen in the main text. This therefore shows that the output angle is tuned by the rotation speed, and the intensity of the beam since $\Delta y:=\Delta y(\Omega,I)$. We also study the transverse shift at different z-positions in the nonlinear and highly nonlinear cases, seen in Fig. \ref{Figure2S}. The transverse shift grows along the direction of propagation. The spacing between curves is non-uniform for different rotation speeds is a clear indicator that the output angle is changed by the nonlinear response of the medium. The transverse shift is plotted in Fig. \ref{Figure2S} for powers of $P_0 = 100$, and $P_0 = 520$ mW. We exclude the linear regime $P_0 = 0.2$ mW since it is on the order of the system noise. One can see that in the limits of high or low rotation speed, the amount of transverse shift is roughly equal in magnitude. It is only in the region mid-range of speeds where the spacing is very non-uniform. One can represent the data in a different way by plotting the amount of transverse shift versus z-position. This can help understand what is going on inside the crystal where imaging of nonlinear systems creates ambiguity in the measurements. One could extrapolate the function back to the crystal front face $z=-20$ mm and see the trajectory along z. For the case of  $\Omega = 100$ deg/s, using linear regression shows a non-zero value at the crystal front face showing evidence that the beam propagation is deviating a straight-line and could be curved based on the nonlinear response of the system. Figures \ref{Figure3aS}, \ref{Figure3S}\textbf{(a)}, and \ref{Figure3S}\textbf{(b)} show the extrapolation of the transverse shift along the z-direction. The progression of these beams should follow a linear regression as the drag is calculated taking an average position of the COM trajectories, which follows a straight line. Instead, as stated before, taking a linear regression gives non-zero values for certain rotation speeds, and thus the trajectory of the beam along z could be curved, or deflected due to a moving index gradient via the nonlinear refraction.

\begin{figure}[t!]                   %Fig9
\centering
\includegraphics[width=0.97\linewidth]{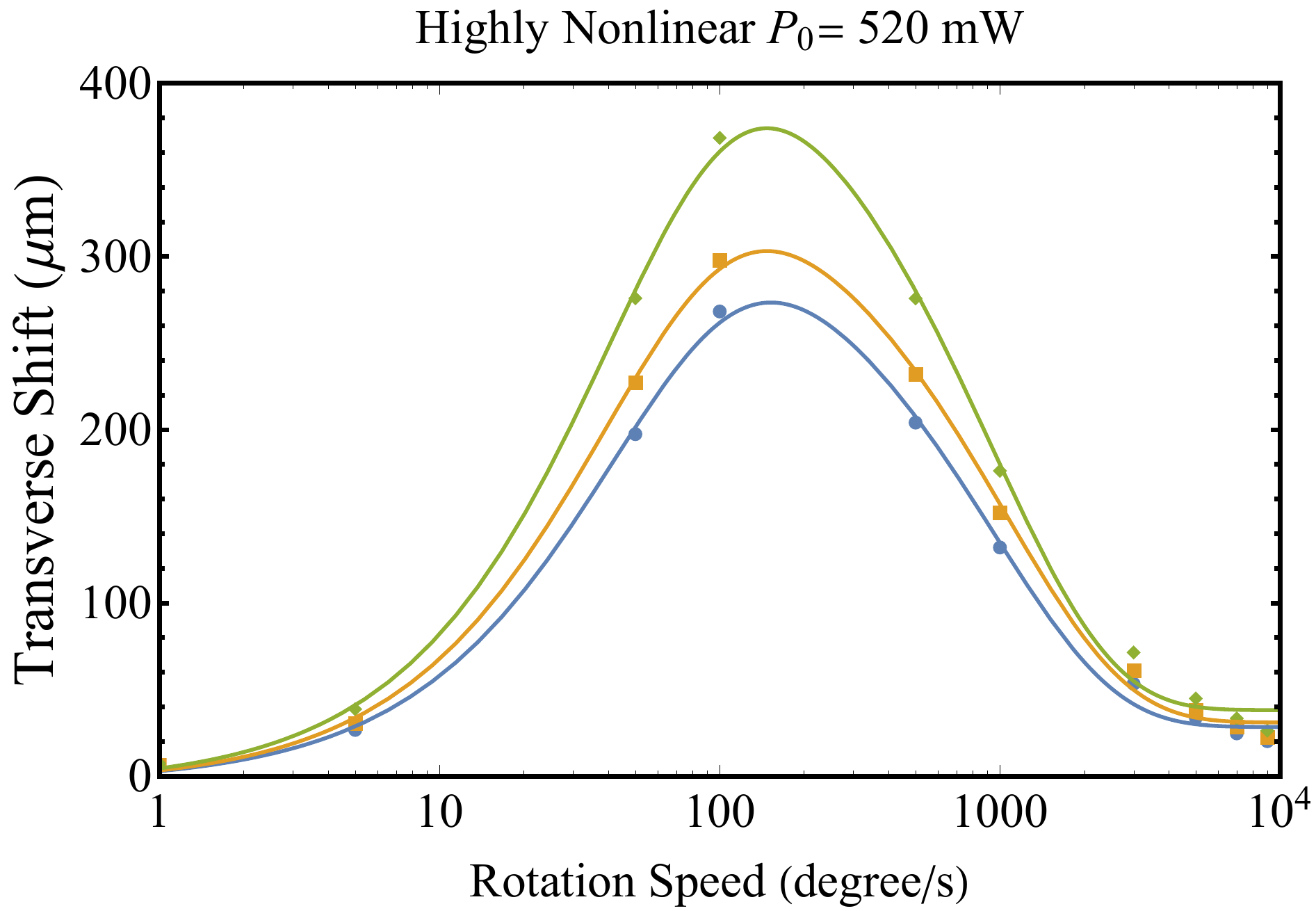}
\caption{A continuous fitting function consisting of the sum of two exponentials. The transverse shift is plotted for the highly nonlinear regime $P_0 = 520$ mW for three positions along z: $z=0$ in blue, $z=7.62$ mm, in yellow, and $z=15.24$ mm in green. Similar behaviour can be seen in the case of an input power of $P_0 = 100$ mW with transverse shifts of smaller magnitude. The form of the fitting function is $f(\Omega) = a+b e^{-\Omega/c}+d e^{ -\Omega/f}$, where a, b, c, d, and f are fitting constants. One can see that the maximum drag should be closer to $\Omega_c\approx 150$ deg/s, while discrete points in Fig. \ref{Figure3S} suggest 100 degs/s. It is clear that the two exponentials indeed fit the discrete points from low to high rotation speeds and provides strong evidence that the transverse shift scales with a sum of exponentials.}
\label{Figure4S} 
\end{figure}

Figure (\ref{Figure3aS}) shows the transverse shift along z for various rotation speeds from 50 to 9000 deg/s in the linear regime ($P_0 = 0.2$ mW). The traces along z for different rotation speeds agrees with the results of the main text. There is indeed no discernable behaviour that can be extracted and the variation is on the order of the system noise. amount of transverse shift in the highly nonlinear regime ($P_0 = 520$ mW). We indeed see the linear behaviour in the highly nonlinear regime is consistent with the COM travelling in a straight line. The magnitudes of the output angle are much larger due to the larger nonlinear response within the crystal. Although the curves slightly deviate from straight lines, this can be attributed to the measurement error in our system.

\subsection*{Fitting function for effective group index}
We create a continuous fitting function consisting of the sum of two exponentials, and an offset constant to fit the transverse shift. The transverse shift is plotted in Fig. \ref{Figure4S} for the highly nonlinear regime $P_0 = 520$ mW for three positions along z: $z=0$ in blue, $z=7.62$ mm, in yellow, and $z=15.24$ mm in green. Plots for the nonlinear regime ($P_0 = 100$ mW) are not shown here but follow similar behaviour of smaller magnitude. The form of the fitting function is $f(\Omega) = a+b e^{-\Omega/c}+d e^{ \Omega/f}$, where a, b, c, d, and f are fitting constants. In the case of creating a continuous function, one can see that the maximum drag should be closer to $\Omega_c\approx 150$ deg/s, rather than the discrete points that suggest 100 degs/s. It is clear that the two exponentials indeed fit the discrete points taken at low to high rotation speeds and provides strong evidence that the transverse shift scales exponentially. Table \ref{tab:nonlinear} shows the fit parameters for each position along z for an input power of 520 mW. We see that the behaviour follows two exponentials with an offset constant value fits well with our data. This is then used with the simulations parameters to understand  the full system and the nonlinear propagation within and after the crystal. The fitting function acts on a higher order component of the generalized nonlinear Schrodinger equation seen in the main text, manipulating the effective group index.

\subsection*{Townes profile formation and nonlinear refraction estimation}
\begin{figure*}
  \includegraphics[width=0.23\linewidth]{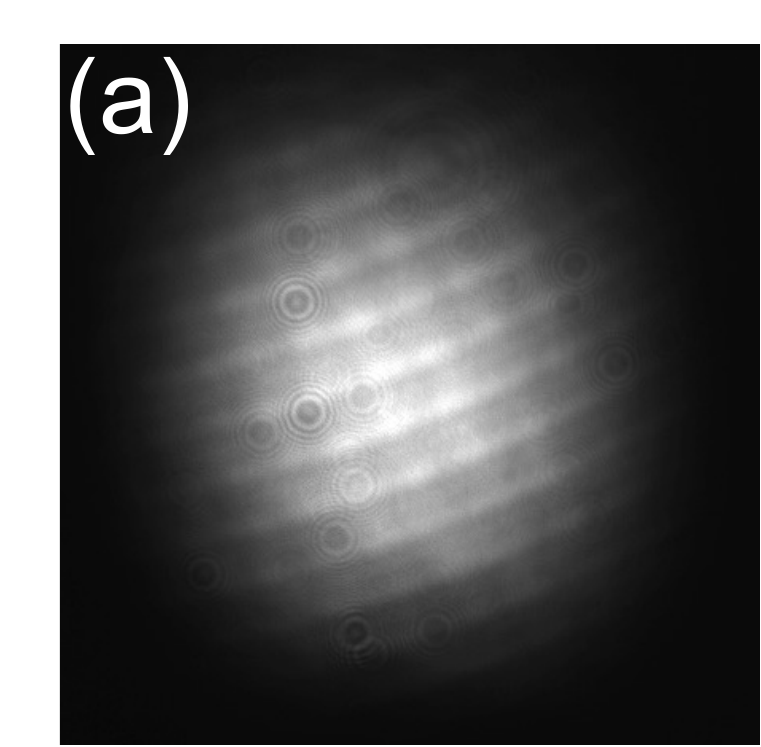}
%   \caption{$P_0 = 10$ mW}
  \includegraphics[width=0.23\linewidth]{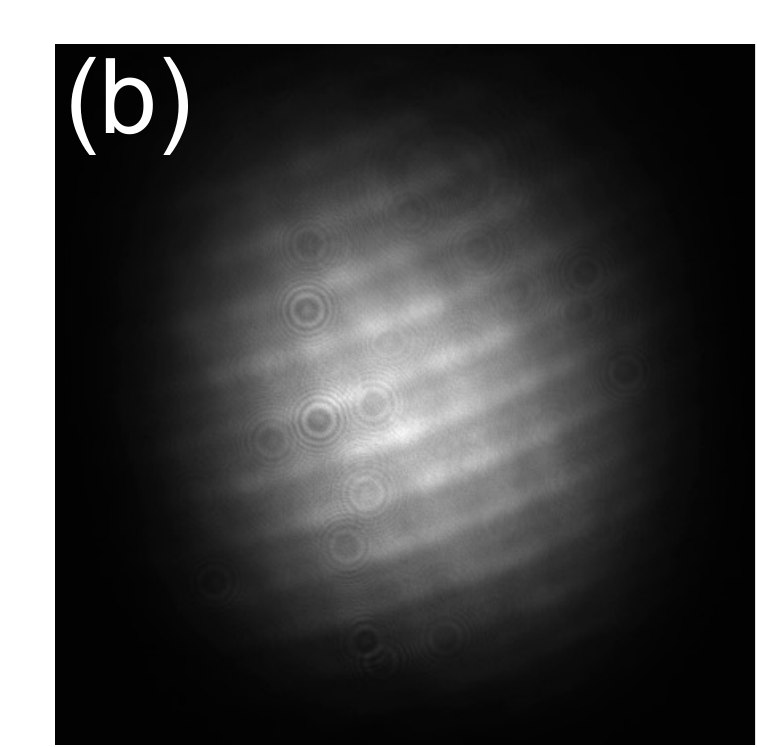}
%   \caption{$P_0 = 12$ mW}
  \includegraphics[width=0.23\linewidth]{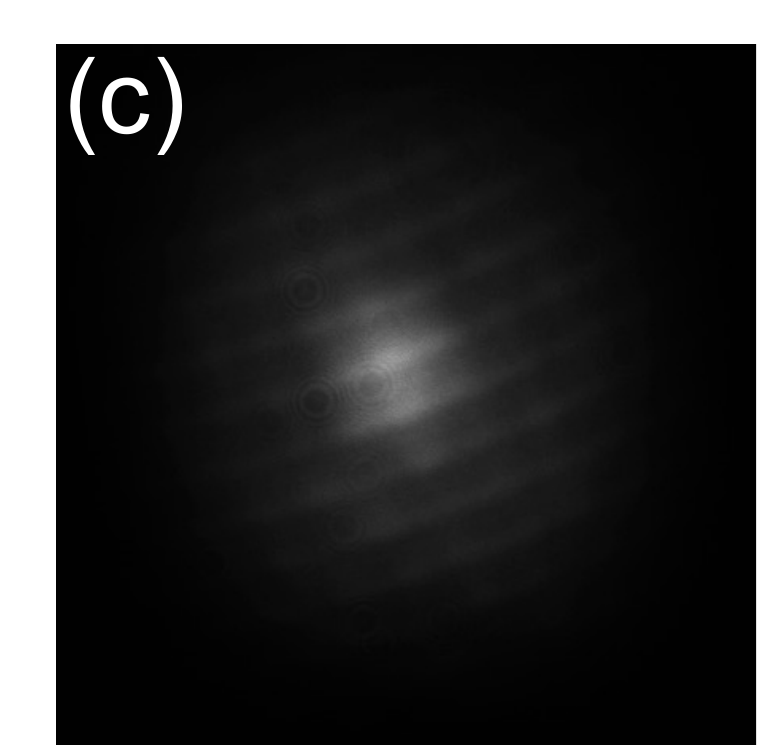}
%   \caption{$P_0 = 398$ mW}
  \includegraphics[width=0.23\linewidth]{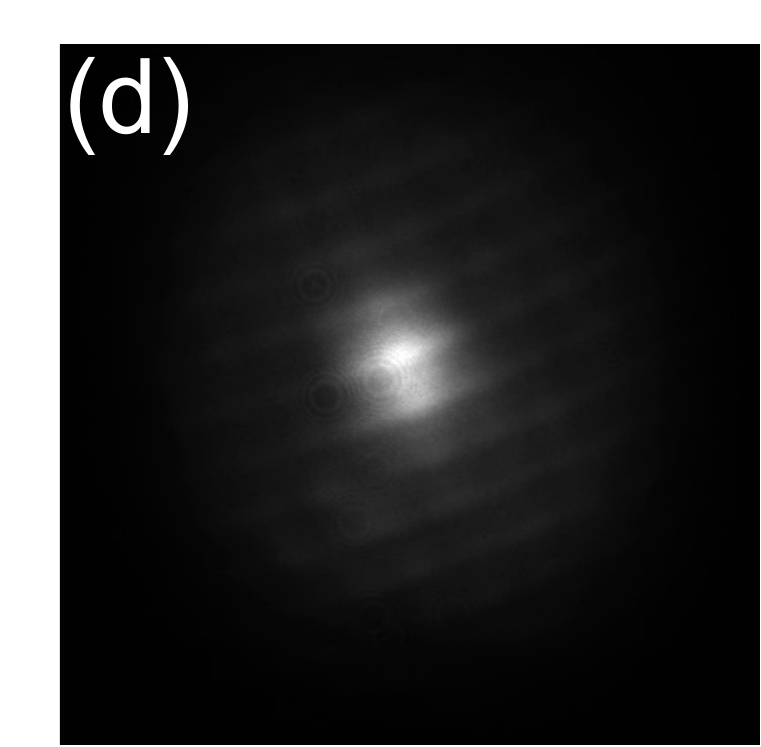}
%   \caption{$P_0 = 520$ mW}
\newline
  \includegraphics[width=0.95\linewidth]{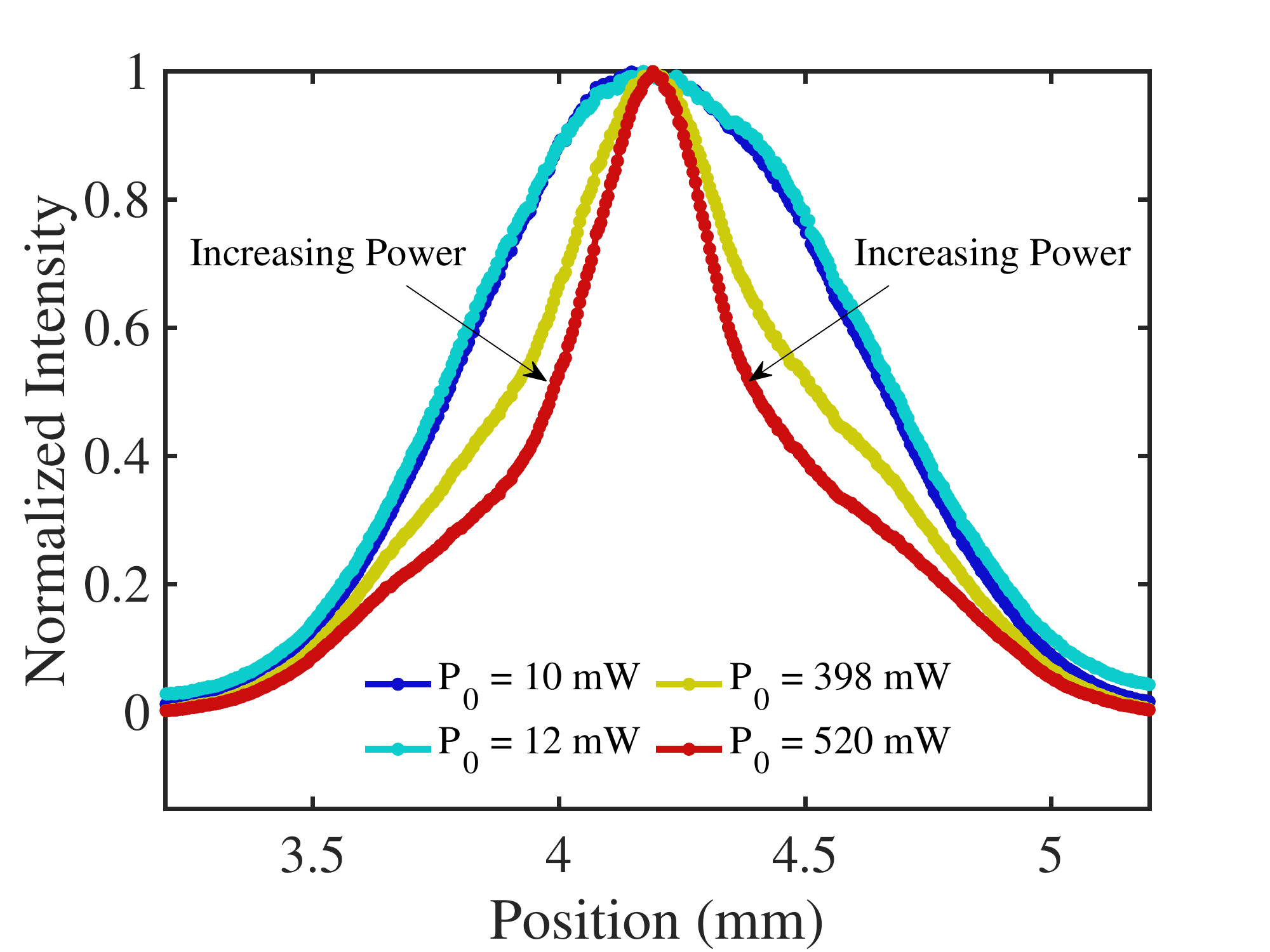}
\caption{The effect of input beam power on beam shape in a stationary medium for a beam waist of 3 mm. Four powers are shown \textbf{(a)} $P_0 = 10$ mW, \textbf{(b)} $P_0 = 12$ mW, \textbf{(c)} $P_0 = 398$ mW, and \textbf{(d)} $P_0 = 520$ mW, where the blue ($P_0 = 10$ mW) and cyan ($P_0 = 12$ mW) curves do not meet the threshold power to show nonlinear refraction and thus self-focusing. Increasing input laser power causes the input beam to self-interact and self-focus creating a spatial soliton. This solitonic behaviour is seen in the yellow ($P_0 = 398$ mW) and red ($P_0 = 520$ mW) curves which are significantly more intense and show a change to the beam's transverse profile. One can see that moderate intensity ($P_0 = 398$ mW) shows slightly less self-focusing than that of the red curve ($P_0 = 520$ mW). The red curve approaches a stable solitonic type solution, known as the Townes Profile. The tapering and stabilization of the beam waist for a Gaussian beam as a result of a self-focusing nonlinearity is a well-known characteristic of spatial solitons. The observation of the Townes profile here indicates there is a considerably large nonlinear index in the system at an input wavelength of $\lambda_0=473$ nm. The beam is not focused by a lens in this case and is the straight output of the laser with a beam diameter of 3 mm. Townes profile formation with CW lasers is uncommon as most soliton solutions are formed using pulsed lasers that needs sufficiently long distances of propagation to stabilize \cite{moll2003self}.}
\label{Figure6S}
\end{figure*}

Although the system is considered instantaneous in this work, it is of interest to study the time response of the system and the ability to reach steady-state. When the ruby crystal is stationary, i,e. $\omega_{rot} = 0$ deg/s, the time response of the system can stabilize and reach steady-state where interesting solutions to wave-propagation are possible. A self-focusing nonlinearity was observed in a stationary version of this experiment, and thus nonlinear refraction contributes to transverse shifting of the beam due to nonlinear deflection. The index gradient impinged on the crystal would move with the rotation of the crystal causing the beam to follow in the direction of higher index. The magnitude of this gradient is
\begin{equation}
\begin{aligned}
 &\Delta n\approx n_2 \frac{2P}{\pi w_o^2} \\&= (10^{-12} m^2/W)\frac{2\times(520\times10^{-3} W)}{\pi (10^{-5} m)^2} = 3.3\times10^{-7},
 \end{aligned}
\end{equation} 
where $n_2\approx10^{-8} m^2/W$ is the nonlinear refractive index, $P$ is the power, and $w_o$ is the beam waist. The value of $n_2$ is not known at 473 nm, however the value was measured at 532 nm and used in this calculation as an order of magnitude estimation \cite{lee1990measurements}. Fig. \ref{Figure6S} shows evidence to formation of a well-known soliton solution, the Townes Profile. This supports the fact that there is a self-focusing nonlinearity in our system, and further studies are needed to quantify the value of $n_2(\lambda = 473 \text{nm})$ and fit the Townes profile. 

\begin{table}[t!]
\begin{tabular}{|c|c|c|c|c|c|}
\hline\hline
& \multicolumn{5}{|c|}{Fitting Parameters}\\
\hline 
{Position (mm)} & $a$ & $b$ & $c$ & $d$ & $f$\\
\hline 
0 & 28.4 & 334 & 46.3 & 302 & 959\\
\hline 
7.52 & 31.0 & 362 & 42.7 & 326 & 1052\\
\hline 
15.24 & 38.1 & 456 & 44.7 & 412.8 & 937\\
\hline\hline 
\end{tabular}
\caption{Results of the fitting parameters  $a$, $b$, $c$, $d$, $f$, $g$ and $h$ for the effective group index in the highly nonlinear ($P_0 = 520$ mW) regime for all rotation speeds ($\Omega = 1-9000$ deg/s). We fit a continuous function consisting of two exponentials and a constant offset of the form: $f(\Omega) = a+b e^{-\Omega/c}+d e^{ -\Omega/f}$.}
\label{tab:nonlinear}
\end{table}
\noindent 

The beam waist used here was much bigger than that of the one used in the transverse drag experimental data, and thus the timescale to reach equilibrium is much longer. Fig. \ref{Figure6S} suggests that nonlinear refraction is strong enough to match the amount of diffraction in the system and stabilize into a steady-state solution. Here, the change in waist should be equivalent to the change in index due to the nonlinear refraction. Further work is needed to measure these values, but it is clear that the self-focusing length scale is on the order of the diffraction length scale. Furthermore, the non-instantaneous response of the system will be the subject of a further study to understand how the timescale can affect the amount of transverse shift. Furthermore, once the beam reaches steady-state, the transverse beam profile reaches a form of the Townes profile. Since the system is using a continuous-wave laser passing through a 2-cm-long solid-state ruby rod,  forming a soliton is interesting as most solitons are created using pulsed lasers that propagate over several metres. Moreover, it would be of interest to see if solitons are also subject to drag and can propagate without breaking up. 
\newpage

\end{document}